\documentclass[acmsmall]{acmart}

\AtBeginDocument{%
  \providecommand\BibTeX{{%
    \normalfont B\kern-0.5em{\scshape i\kern-0.25em b}\kern-0.8em\TeX}}}

\usepackage{framed}
\usepackage{multirow}
\usepackage{booktabs}
\usepackage{ifthen}
\usepackage{color}
\usepackage{url}
\usepackage{amsmath}
\usepackage{xspace}
\usepackage[T1]{fontenc}
\usepackage{enumerate}
\usepackage[linesnumbered,ruled,vlined]{algorithm2e}
\usepackage{array,multirow,graphicx}
\usepackage{float}
\usepackage{balance}
\usepackage{tikz}
\usepackage{calc}
\usepackage{subfigure}
\usepackage{listings,amsfonts}
\usepackage{amsmath,xcolor,pifont}
\usepackage{url}
\usepackage{xspace}
\usepackage{hyperref,endnotes}
\usepackage{xcolor}
\usepackage{array}
\usepackage{multirow,makecell}
\usepackage[misc]{ifsym}
\usepackage[toc,title,page]{appendix}
\usepackage{mathrsfs}

\definecolor{yellow}{RGB}{255,255,153}
\definecolor{grey}{RGB}{224,224,224}

\begin{document}

\setcopyright{acmcopyright}
\acmJournal{POMACS}
\acmYear{2020} \acmVolume{4} \acmNumber{3} \acmArticle{50} \acmMonth{12} \acmPrice{15.00}\acmDOI{10.1145/3428335}
\received{August 2020}
\received[revised]{September 2020}
\received[accepted]{October 2020}

\title{Tracking Counterfeit Cryptocurrency End-to-end}

\author{Bingyu Gao}
\affiliation{%
  \institution{Beijing University of Posts and Telecommunications}
  \city{Beijing}
  \country{China}
}

\author{Haoyu Wang}
\authornote{Corresponding Author: Haoyu Wang (haoyuwang@bupt.edu.cn).}
\affiliation{%
  \institution{Beijing University of Posts and Telecommunications}
  \city{Beijing}
  \country{China}
}
\email{haoyuwang@bupt.edu.cn}

\author{Pengcheng Xia}
\affiliation{%
  \institution{Beijing University of Posts and Telecommunications}
  \city{Beijing}
  \country{China}
}

\author{Siwei Wu}
\affiliation{%
  \institution{Zhejiang University}
  \city{Hangzhou}
  \country{China}
}

\author{Yajin Zhou}
\affiliation{%
  \institution{Zhejiang University}
  \city{Hangzhou}
  \country{China}
}

\author{Xiapu Luo}
\affiliation{%
  \institution{The Hong Kong Polytechnic University}
  \city{HongKong}
  \country{China}
}

\author{Gareth Tyson}
\affiliation{%
  \institution{Queen Mary University of London}
  \city{London}
  \country{United Kingdom}
}

\begin{abstract}
The production of counterfeit money has a long history. It refers to the creation of imitation currency that is produced without the legal sanction of government.
With the growth of the cryptocurrency ecosystem, there is expanding evidence that counterfeit cryptocurrency has also appeared.
In this paper, we empirically explore the presence of counterfeit cryptocurrencies on Ethereum and measure their impact.
By analyzing over 190K ERC-20 tokens (or cryptocurrencies) on Ethereum, we have identified $2,117$ counterfeit tokens that target 94 of the 100 most popular cryptocurrencies. We perform an end-to-end
characterization of the counterfeit token ecosystem, including their popularity, creators and holders, fraudulent behaviors and advertising channels. 
Through this, we have identified two types of scams related to counterfeit tokens and devised techniques to identify such scams. We observe that over 7,104 victims were deceived in these scams, and the overall financial loss sums to a minimum of \$ 17 million (74,271.7 ETH). Our findings demonstrate the urgency to identify counterfeit cryptocurrencies and mitigate this threat. 

\end{abstract}



\maketitle

\section{Introduction}

Since the first Bitcoin block was mined in 2009, cryptocurrencies have seen significant growth. This growth is mainly due to the rapid development of blockchain technologies and the digital economic system. Besides Bitcoin, thousands of cryptocurrencies have emerged. As of the end of 2019, the total market capitalization of cryptocurrencies is over \$180 billion~\cite{cryptocurrenciesmarket2019}.

\textit{Where there is money, there are those who follow it.} Cryptocurrencies have attracted extensive attention from attackers. 
Attackers have exploited the vulnerabilities in smart contracts, cryptocurrency exchanges and wallets.
According to endpoint security provider Carbon Black, \$1.1 billion in cryptocurrency was stolen in attacks during the first half of 2018~\cite{cryptocurrencystolen}. As reported in May 2019, attackers have stolen 7,000 bitcoins (worth \$41m) from Binance, one of the top leading exchanges~\cite{binancehackers}.
Hundreds of popular Gambling Decentralized Applications (DApps), Defi Dapps, and other smart contracts were attacked recently, causing huge economic losses.
Besides these known attacks, a number of newly emerging scams are taking advantage of cryptocurrencies to make a profit. 
For example, Marie Vasek and Tyler Moore~\cite{vasek2015there} presented the first empirical analysis of Bitcoin-based scams in 2015, including high-yield investment programs, mining investment scams, scam wallet services and scam exchanges.
After that, some other studies have characterized various scams including cryptocurrency Ponzi Schemes~\cite{vasek2018analyzing, bartoletti2020dissecting, chen2018detecting}, blockchain honeypots~\cite{torres2019art}, extortion emails~\cite{SpamsSextortion}, and cryptocurrency exchange phishing scams~\cite{xia2020characterizing}, etc.

However, an understudied attack is \textit{counterfeit money} --- the imitation of currency, which is produced without the legal sanction of the government~\cite{wiki-counterfeit}. Fraudsters strive to imitate the official currency so as to deceive its recipient.
Following the same postulation, we ask \textit{if such kinds of counterfeited money have appeared in the cryptocurrency ecosystem?}
Evidence suggests that the answer is yes, with several news reports discussing cryptocurrency abuse~\cite{libra-fake, fakehuobinews}. 
For example, since the Libra currency~\cite{libra} was announced, scammers have designed fraudulent investment schemes involving the sale of fake ``Libra tokens'' that are unaffiliated with the actual Libra brand~\cite{libra-fake}.
Similarly, recently, Chinese authorities have seized cryptocurrencies worth over \$10 million while bringing down a scam involving fake HuobiTokens~\cite{fakehuobinews}\footnote{HuobiToken is an ERC-20 token on Ethereum, which is released by Huobi, one of the most popular exchanges.}.

\begin{figure} [t]
\centering
  \includegraphics[width=0.5\textwidth]{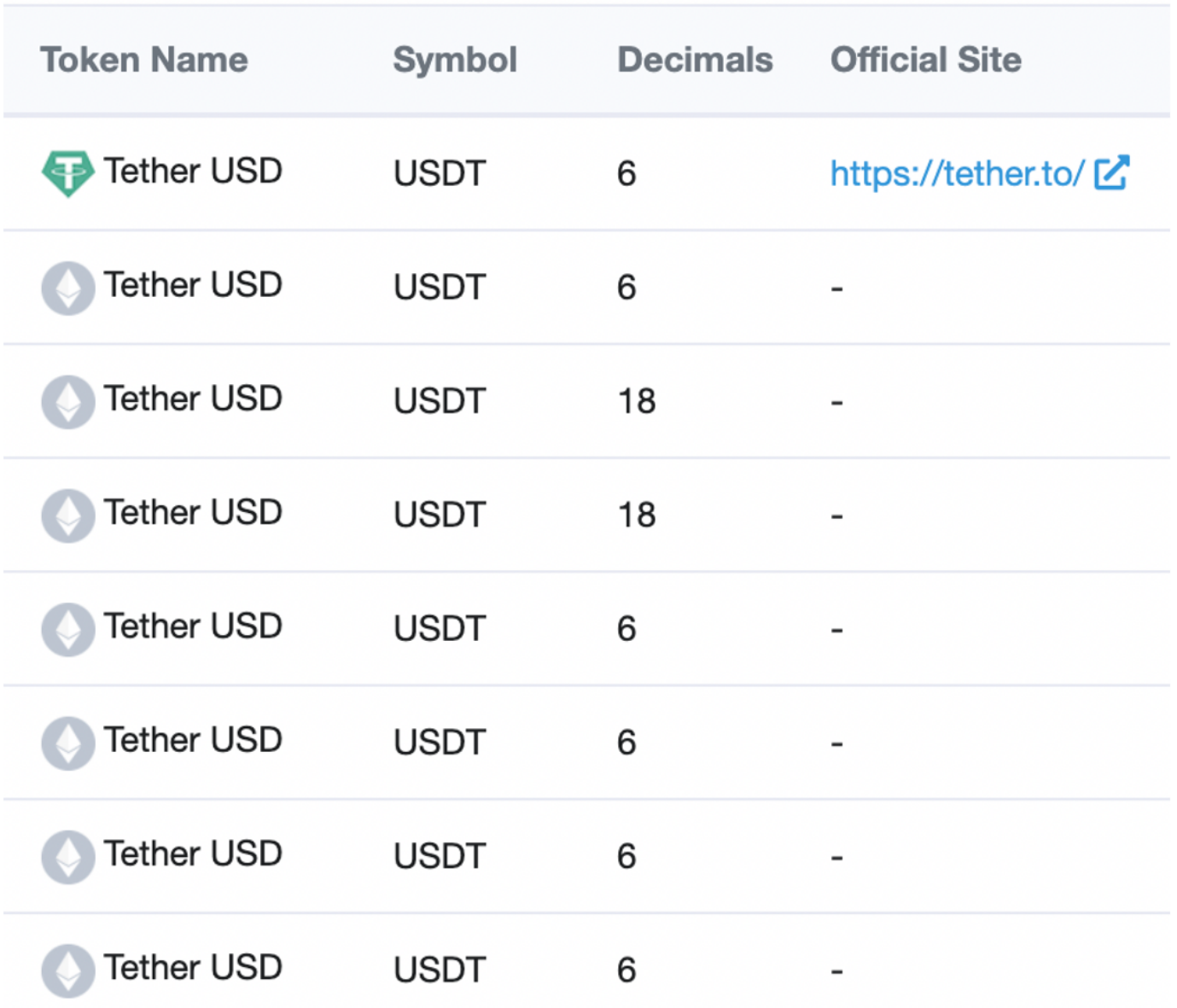}
\caption{A number of ERC-20 tokens with the identical name and symbol of Tether USD (USDT).}
\label{fig:usdt_example} 
\end{figure}

Moreover, the ease of creating cryptocurrencies and launching Initial Coin Offerings (ICO), makes the cost of releasing counterfeit cryptocurrencies quite low. 
For example, Ethereum, as an open-source platform for decentralized applications (DApps), is the first blockchain platform that simplifies the development of smart contracts. Based on Ethereum, one can create a token smart contract with just a few lines of code.
By July 2020, there were over 200 thousand ERC-20 tokens created on Ethereum~\cite{EtherscanERC20}.
However, Ethereum does not enforce any restrictions on the names and symbols of the newly created tokens. Instead, the thing that identifies a token is its smart contract address.
As shown in Figure~\ref{fig:usdt_example}, by searching Tether USD (USDT), a popular token that attempts to be tied to the US dollar, in the Etherscan (the most widely used Ethereum explorer)~\cite{EtherscanERC20}, there are over 170 tokens with the identical name ``Tether USD'' or symbol ``USDT''. \textit{This opens up a number of potential fraudulent avenues, with malicious parties potentially exploiting this fact to counterfeit cryptocurrencies.} 

Despite this, to the best of our knowledge, the counterfeit cryptocurrency ecosystem has not been systematically investigated or measured. 
Thus, there is a general lack of an understanding of this attack, including: 1) \textit{to what extent counterfeit cryptocurrencies exist}; 2) \textit{what are the entities related to the counterfeit cryptocurrencies}, i.e., their targets, creators, distributors, and users; 
3) \textit{what are they used for}, i.e., whether counterfeit cryptocurrencies are involved in blockchain scams; 
and 4) \textit{the advertising channels of counterfeit cryptocurrencies}, i.e., how do they reach and attract victims.

\textbf{This Work.} 
In this paper, we present the first systematic study of counterfeit cryptocurrencies on Ethereum. 
By analyzing over 170K ERC-20 tokens created on Ethereum (before March 2020), we have identified 2,117 counterfeit tokens that target 94 of the top-100 most popular cryptocurrencies (tokens). 
We then analyze the distribution and popularity of these counterfeit tokens, as well as the creators and holders of them (see \textbf{Section~\ref{sec:measurement}}).
After identifying two types of fraudulent behaviors related to the counterfeit cryptocurrencies, we further measure the impacts of the counterfeit cryptocurrency ecosystem, including the scale of the financial losses and the number of victims (see \textbf{Section~\ref{sec:scam}}).
Finally, to further understand how they are spread, we go a further step to investigate the advertising channels of these counterfeit cryptocurrencies (see \textbf{Section~\ref{sec:advertising}}).
This paper reveals the ecosystem of counterfeit cryptocurrencies with some unexpected and interesting observations:

\begin{itemize}
    \item \textbf{Counterfeit tokens are prevalent on Ethereum.} 
    94\% (94/100) of the official tokens we studied in this paper have already been targeted by counterfeit tokens. Some counterfeit tokens are quite popular, with thousands of transactions and holders.
    
    \item \textbf{The scams related to counterfeit tokens cause huge financial losses.} 
    We have characterized two types of scams related to counterfeit tokens, and quantified the direct financial impact. The overall volume is a minimum amount of \$17 million.
    
    \item \textbf{A number of reputable platforms are abused to help spread the fraudulent information of counterfeit tokens.} We have identified 935 pieces of advertising information related to counterfeit tokens from 103 well-known platforms. These include Telegram, Facebook, Bitcointalk (the official forum of Bitcoin), YouTube, etc. Various kinds of social engineering techniques are abused by attackers to attract victims.
\end{itemize}

\section{Background}
\label{sec:background}

\subsection{Blockchain and Ethereum}
Blockchain, which was invented in 2008 by Satoshi Nakamoto, is an open distributed ledger that stores transactions or related events among involved parties. It is maintained by a peer-to-peer network and secured by cryptographic design, thus it is resistant to data modification. By this design, each transaction in the block is verified by the confirmation of most participants in the system. Blockchain was originally served as a ledger for the Bitcoin, the first decentralized cryptocurrency. Bitcoin demonstrated
the feasibility to construct a decentralized value-transfer system that can be shared across the world and virtually free to use.
Bitcoin's blockchain design has inspired many other blockchain systems like Ethereum and EOSIO. 

Ethereum is an open-source decentralized blockchain platform featuring smart contract functionality. It was proposed by Vitalik Buterin, and its development was funded by an online crowdsale. After that, it was initially released in 2015. \textit{Ether} (ETH) is the cryptocurrency mined by Ethereum miners as a reward for computations and it is the second largest cryptocurrency based on volume.

\subsection{Ethereum Account and Transactions}

\textbf{Ethereum Account.}
An account is the basic unit to identify an entity in Ethereum. An account is identified by a fixed-length hash-like address.
Ethereum has two kinds of accounts: \textit{external owned accounts} (EOAs) that are controlled by public-private key pairs (i.e. humans); and \textit{contract owned accounts} (COAs) controlled by the code stored together with the account. 
An EOA is an ordinary account that can transfer tokens, invoke deployed smart contracts and store received tokens. Moreover, an EOA can deploy a smart contract into a COA account.
All accounts are referred to by their addresses and denoted as an six-character identifier beginning with \texttt{0x} in this paper.

\textbf{Transaction.}
A transaction in Ethereum is a message sent from one account to another, which records the state changes of accounts. ``Gas'', an internal transaction pricing mechanism, is used to protect the blockchain from spam and allocate resources on the network during transactions.
A transaction can include binary data (called the ``payload'') and Ether. There are two kinds of transactions depending on the message sender. The transactions sent from an EOA are called ``external transactions'', which will be included in the blockchain and can be obtained by parsing the blocks. The other type, initiated by executing a smart contract, is called ``internal transaction''. Internal transactions are usually triggered by external transactions and are not stored in the blockchain directly.

\subsection{Smart Contract and ERC-20 Token}
\textbf{Smart Contract.}
Smart contracts are a kind of decentralized agreement built by computer programs, which are used to implement arbitrary rules as well as guaranteeing to produce the same result for decentralized parties.
In Ethereum, a smart contract is a collection of code and data that reside in a Contract Account.
A smart contract can be executed automatically, and it can control related events based on the terms built into its code. In the Ethereum platform, it is easy for people to build decentralized apps (DApps) and issue tokens for DApps or other purposes through smart contracts.
Smart contracts are typically written in higher level languages (e.g., Solidity) then compiled to Ethereum Virtual Machine (EVM) bytecode. EVM is the runtime environment for smart contracts in Ethereum. 

\textbf{Token.}
In contrast to digital coins like Bitcoin and Ether, which are native to their own blockchain, tokens require existing blockchain platforms. Based on the function of tokens, they are usually classified into three types~\cite{tokentype}: 1) \textit{currency tokens}, which are entirely created as a method of payment; 2) \textit{utility tokens}, which grant investors access to some kinds of products or services; and 3) \textit{investment/asset tokens}, which are the assets that promise investors a return on their investment. The most famous investment token is the decentralized autonomous organization (DAO) token. It is an ERC-20 token which was a form of investor-directed venture capital fund, and whose vulnerabilities led to a hard fork of Ethereum.
Note that every token that exists on the Ethereum is tied to a token contract, which defines a set of functions they use to perform tasks.

\begin{figure} [t]
\centering
\includegraphics[width=0.6\textwidth]{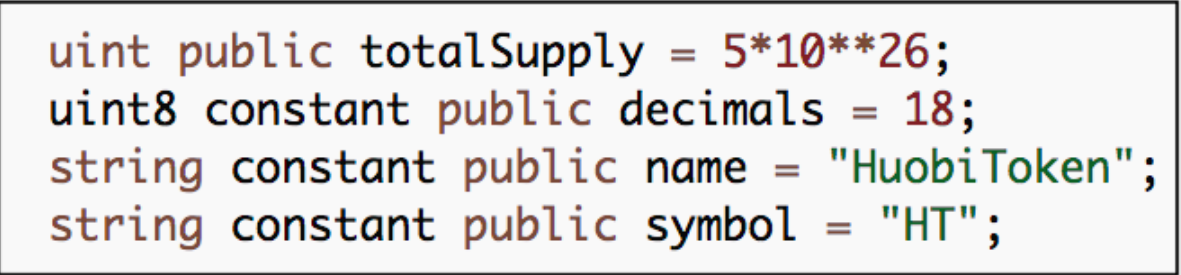}
\caption{The code snippet of an ERC-20 token smart contract.}
\label{fig:tokenexample} 
\end{figure}

\textbf{ERC-20.}
ERCs (Ethereum Request for Comments) are technical documents used by smart contract developers at Ethereum, which define a set of rules required to implement tokens for the Ethereum ecosystem.
ERC-20 is by far the most recognizable token standard.
It is proposed for developers to better handle different tokens on Ethereum. 
An ERC-20 token contract usually has properties including name, symbol, total supply and decimal, etc. 
An example of an ERC-20 token is shown in Figure~\ref{fig:tokenexample}, whose token name is ``HuobiToken'' and symbol is ``HT''. 
Due to the ERC-20 standard, Ethereum has become one of the most popular token platforms --- as of July 2020, there are over 200,000 ERC-20 tokens on Ethereum. Note that, in this paper, each token is represented in the form of \texttt{TokenName (SymbolName)}.

\subsection{Counterfeit Cryptocurrency}
\label{subsec:definition}

Ethereum does not enforce any restrictions on the names and symbols of newly created tokens, even if the names have been used by existing tokens.
\textit{This, however, could be abused by attackers to create counterfeit cryptocurrencies}.
Just like producing counterfeit money by imitating fiat money (e.g., US dollars), attackers may use same identifiers (e.g., token name and symbol) or confusingly similar identifier names to deceive inexperienced investors. 
Thus, we consider the following two types of counterfeit tokens in this paper:
\begin{itemize}
    \item \textbf{Type-1} The counterfeit token has an identical identifier name to an imitated official cryptocurrencies, but is released by different creators.
    
    \item \textbf{Type-2} The attacker adopts combo-squatting techniques~\cite{combosquatting}\footnote{Combo-squatting is a specific type of domain squatting, in which attackers register domains that combine a recognizable brand name (e.g., Paypal) with other keywords (e.g., login). Combo-squatting attacks are prevalent in domain~\cite{combosquatting} and even mobile apps~\cite{hu2020mobile}.} to create a counterfeit token. This involves the combination of a recognizable token name (e.g., USDT) with other characters or keywords (e.g., USDT-2, USDT New). Such counterfeit tokens often have confusingly similar names to official cryptocurrencies, leading people to believe they are the new version of the official tokens or at least released by the same team.
\end{itemize}

\section{Study Design}
\label{sec:design}

We present the details of our characterization study on counterfeit cryptocurrencies in this section.
We first describe our research questions, and then present the dataset used for our study. Last, we discuss our rationale for selecting the tokens that are most likely to be abused.

\subsection{Research Questions}
\label{subsec:RQ}

Our study aims to investigate the overall ecosystem of counterfeit cryptocurrencies, from their creation, transaction and circulation, to the scams related to them and their advertising channels.
To this end, our study is driven by the following research questions (RQs):

\begin{itemize}
    \item[RQ1] \textbf{\textit{Are counterfeit cryptocurrencies prevalent in the cryptocurrency ecosystem?}} 
    Although a few reports in the media mentioned the existence of counterfeit cryptocurrencies, their scale remains unknown. Furthermore, it is necessary to investigate: 
    RQ1.1) \textit{Which tokens are their targets?} Considering there are thousands of cryptocurrencies, we seek to explore whether adversaries predominantly target tokens with greater popularity (volume).
    RQ1.2) \textit{who create counterfeit cryptocurrencies?} It is interesting to study whether some malicious campaigns habitually create a number of counterfeit cryptocurrencies to deceive unsuspecting users or investors.
    RQ1.3) \textit{who hold these counterfeit cryptocurrencies?}
    Analyzing the holders of these counterfeit tokens can help us understand the overall scale of the ecosystem, i.e., how many accounts have been involved in counterfeit tokens. RQ1 will be studied in Section~\ref{sec:measurement}.
    
    \item[RQ2] \textbf{\textit{What are the fraudulent behaviors related to counterfeit cryptocurrencies?}} 
    We are still unaware of the usage of the counterfeit tokens. Thus, it is interesting to investigate how users can be scammed by counterfeit cryptocurrencies, whether there are other collusion addresses and even malicious campaigns involved in the scams, and how many users were scammed by them? RQ2 will be studied in Section~\ref{sec:scam}.

    \item[RQ3] \textbf{\textit{What are the advertisement channels of counterfeit cryptocurrencies?}} 
    It is interesting to analyze how counterfeit currencies reach users, especially the advertising channels, tricks and social engineering techniques adopted. This can help us better identify and trace scams and malicious campaigns behind. RQ3 will be studied in Section~\ref{sec:advertising}.
\end{itemize}

\subsection{Datasets}
\label{subsec:datasets}

Since our goal is to measure counterfeit cryptocurrencies in Ethereum, we require both 1) a complete list of the \textit{ERC-20 tokens}, which is used for detecting counterfeit tokens; and 2) the whole \textit{Ethereum transaction dataset}, which is used to analyze transactions related to counterfeit cryptocurrencies.

Thus, we take advantage of Geth\footnote{https://geth.ethereum.org/}, a widely-used Ethereum client to synchronize the ledger of Ethereum.
We have synchronized all the blocks until March 18th, 2020, with over 9.6 million blocks in total.
The data extracted from the blocks contains \textit{external transactions}, \textit{internal transactions}, \textit{contract information}, and \textit{contract calling information}. 
We then get the bytecode and creator information for all the smart contracts.
Then, we analyze the bytecode to determine whether a contract implements an ERC-20 token. Based on this method, we have identified over 176K ERC-20 tokens alongside their creator information. We further obtain the metadata of these tokens (e.g., website, total supply, holder, etc.) from either the code or Etherscan.
To facilitate the analysis, we use ElasticSearch to store these structured data, which provides a query interface for our characterization study.

\subsection{Target Cryptocurrency Selection}
\label{subsec:target}

While all the tokens could be the subject of counterfeit cryptocurrencies, it is arguably not in the best interest of an attacker to use a less known token for abuse (e.g., an official token with less user and volume).
Furthermore, as there are thousands of official tokens on Ethereum (mixed with counterfeit tokens), it is hard for us to compile a complete list of all the official tokens. 
We do this to reduce the number of potential counterfeit currencies that must be measured. 
Consequently, we compile a list of the top-100 tokens based on market capitalization from Etherscan.
Table~\ref{tab:targettoken} shows the information of these official tokens. Column 1 (\#CAP) shows the ranking of the tokens based on their market capitalization by the time of our study.

\begin{table}[t]
\centering
\caption{The target official tokens and their counterfeit tokens identified (CTokens). Column 5 (\#Trans) shows the aggregated number of all transactions related to counterfeit tokens for the corresponding official token.}
\vspace{-0.1in}
\label{tab:targettoken}
\resizebox{\linewidth}{!}{
\begin{tabular}{|l|l|l|l|l||l|l|l|l|l|}
\hline
\# CAP & Token Name              & Symbol & \# CTokens & \# Trans & \# CAP & Token Name          & Symbol  & \# CTokens & \# Trans \\ \hline
1  & Tether USD              & USDT    & 171 & 12,188 & 51  & Ampleforth        & AMPL  & 5  & 202  \\\hline
2  & BNB                     & BNB     & 90  & 781    & 52  & Pundi X Token     & NPXS  & 5  & 31   \\\hline
3  & ChainLink Token         & LINK    & 15  & 24     & 53  & Trace             & TRAC  & 6  & 1,262 \\\hline
4  & HuobiToken              & HT      & 545 & 12,883 & 54  & Tellor Tributes   & TRB   & 21 & 89   \\\hline
5  & Bitfinex LEO Token      & LEO     & 20  & 306    & 55  & MXCToken          & MXC   & 0  & 0    \\\hline
6  & Crypto.com Coin         & CRO     & 1   & 0      & 56  & Uquid Coin        & UQC   & 3  & 10   \\\hline
7  & -                       & HEDG    & 5   & 69     & 57  & BandToken         & BAND  & 1  & 6    \\\hline
8  & Maker                   & MKR     & 10  & 102    & 58  & Synth sUSD        & sUSD  & 3  & 61   \\\hline
9  & USD Coin                & USDC    & 70  & 3,303  & 59  & Insolar           & INS   & 1  & 2    \\\hline
10 & OKB                     & OKB     & 60  & 1,451  & 60  & OriginToken       & OGN   & 10 & 202  \\\hline
11 & Ino Coin                & INO     & 1   & 41     & 61  & Melon Token       & MLN   & 8  & 316  \\\hline
12 & VeChain                 & VEN     & 8   & 20     & 62  & Utrust Token      & UTK   & 9  & 68   \\\hline
13 & BAT                     & BAT     & 69  & 277    & 63  & Wrapped BTC       & WBTC  & 8  & 353  \\\hline
14 & Paxos Standard          & PAX     & 21  & 417    & 64  & Rocket Pool       & RPL   & 5  & 14   \\\hline
15 & ZRX                     & ZRX     & 25  & 369    & 65  & Pinakion          & PNK   & 0  & 19   \\\hline
16 & Insight Chain           & INB     & 3   & 3      & 66  & Ankr Network      & ANKR  & 14 & 177  \\\hline
17 & ICON                    & ICX     & 9   & 19     & 67  & QuarkChain Token  & QKC   & 39 & 1,297 \\\hline
18 & OMG Network             & OMG     & 54  & 552    & 68  & DATAcoin          & DATA  & 21 & 554  \\\hline
19 & Republic                & REN     & 7   & 78     & 69  & Chimpion          & BNANA & 1  & 1    \\\hline
20 & Baer Chain              & BRC     & 19  & 6,456  & 70  & Numeraire         & NMR   & 7  & 511  \\\hline
21 & ZBToken                 & ZB      & 13  & 73     & 71  & Reserve Rights    & RSR   & 5  & 9    \\\hline
22 & Synthetix Network Token & SNX     & 17  & 38     & 72  & SingularityNET    & AGI   & 12 & 273  \\\hline
23 & TrueUSD                 & TUSD    & 52  & 573    & 73  & BTU Protocal      & BTU   & 1  & 1    \\\hline
24 & HoloToken               & HOT     & 16  & 298    & 74  & SwissBorg         & CHSB  & 2  & 1    \\\hline
25 & Dai Stablecoin          & DAI     & 69  & 991    & 75  & Paxos Gold        & PAXG  & 4  & 0    \\\hline
26 & Mixin                   & XIN     & 3   & 3      & 76  & Polymath          & POLY  & 45 & 1,973 \\\hline
27 & Theta Token             & THETA   & 10  & 21     & 77  & Fantom Token      & FTM   & 20 & 260  \\\hline
28 & Nexo                    & NEXO    & 12  & 157    & 78  & Ocean Token       & OCEAN & 3  & 713  \\\hline
29 & Cryptonex               & CNX     & 0   & 0      & 79  & Gnosis            & GNO   & 13 & 66   \\\hline
30 & Kucoin Shares           & KCS     & 6   & 12     & 80  & Bancor            & BNT   & 18 & 32   \\\hline
31 & Sai Stablecoin v1.0     & SAI     & 11  & 510    & 81  & Aragon            & ANT   & 5  & 136  \\\hline
32 & Bytom                   & BTM     & 33  & 446    & 82  & HarmonyOne        & ONE   & 40 & 124  \\\hline
33 & EnjinCoin               & ENJ     & 8   & 417    & 83  & Storj             & STORJ & 7  & 17   \\\hline
34 & MCO                     & MCO     & 3   & 29     & 84  & Io TeX Network    & IOTX  & 5  & 14   \\\hline
35 & DGD                     & DGD     & 4   & 132    & 85  & Fetch             & FET   & 11 & 104  \\\hline
36 & IOSToken                & IOST    & 14  & 198    & 86  & STASIS EURS Token & EURS  & 0  & 0    \\\hline
37 & Centrality Token        & CENNZ   & 1   & 0      & 87  & UniBright         & UBT   & 4  & 58   \\\hline
38 & Zilliqa                 & ZIL     & 16  & 378    & 88  & Enigma            & ENG   & 19 & 157  \\\hline
39 & KyberNetwork            & KNC     & 9   & 120    & 89  & Celsius           & CEL   & 3  & 11   \\\hline
40 & WAX Token               & WAX     & 3   & 260    & 90  & LoopringCoin V2   & LRC   & 7  & 15   \\\hline
41 & StatusNetwork           & SNT     & 8   & 42     & 91  & WaykiCoin         & WIC   & 2  & 5    \\\hline
42 & Golem                   & GNT     & 12  & 125    & 92  & PowerLedger       & POWR  & 32 & 994  \\\hline
43 & SeeleToken              & Seele   & 20  & 30     & 93  & EthLend           & LEND  & 6  & 23   \\\hline
44 & NOAHCOIN                & NOAH    & 17  & 76     & 94  & AION              & AION  & 4  & 30   \\\hline
45 & Reputation              & REP     & 16  & 7      & 95  & Matic Token       & MATIC & 19 & 1,643 \\\hline
46 & RLC                     & RLC     & 1   & 0      & 96  & Decentraland      & MANA  & 49 & 168  \\\hline
47 & Cryptoindex 100         & CIX 100 & 0   & 0      & 97  & chiliZ            & CHZ   & 2  & 202  \\\hline
48 & Banker Token            & BNK     & 5   & 4      & 98  & ELF               & ELF   & 7  & 39   \\\hline
49 & Binance USD             & BUSD    & 18  & 225    & 99  & DxChain Token     & DX    & 0  & 0    \\\hline
50 & EXMER FDN.              & EXMR    & 1   & 7      & 100 & Swipe             & SXP   & 4  & 10  \\\hline

\textbf{-}      & \textbf{-}   & \textbf{-}  & \textbf{-}   & \textbf{-}  & \textbf{-}   & \textbf{Sum}  &         & \textbf{2,117}  & \textbf{56,764}    \\ \hline
\end{tabular}
}
\vspace{-0.1in}
\end{table}

\section{Measurement of Counterfeit Cryptocurrencies}
\label{sec:measurement}

\subsection{Detection Method}
According to the previous definition of the counterfeit token (see Section~\ref{subsec:definition}), we adopt a two-step and semi-automated approach to perform accurate detection of counterfeit tokens. 
\textit{The first step is to identify all the possible counterfeit token candidates by keyword matching}. For the selected 100 official tokens, we search their token names and symbol names in the ERC-20 token dataset, to flag tokens that contain such keywords.
However, we find that the keyword matching based method may introduce a number of false positives. 
Thus, \textit{for the second step, we propose to remove false positives} according to the following rules:

\begin{itemize}
    \item[Rule1] \textbf{Migrated tokens.}
    As smart contracts in Ethereum cannot be modified once deployed, some tokens migrate their addresses due to security concerns or new features being introduced. For example, the token HEDG has migrated from 0xb6B6Bd\footnote{0xb6B6Bd3c75c4237089b5ED518A1809C297CC2e6B} to 0x3363D5\footnote{0x3363D570f6DF3c74d486BB8785d3EbFB9E2347D3}, due to new features being introduced. In this case, we manually analyze the 100 selected tokens to verify whether they have migrated addresses and further remove such false positives.
    
    \item[Rule2] \textbf{Official tokens created by trustworthy creators.}
    It is a common practice that, before releasing a new token officially, some creators release some test tokens to check whether their tokens will perform as expected. 
    These attempts will lead to the creation of a few tokens with the same/similar identifier names. 
    Thus, we manually check the creators of the filtered tokens to remove such false positives.
    
   \item[Rule3] \textbf{Official tokens that have similar names with our target tokens}. As the symbol name is usually short (e.g., 3 to 5 characters), it is quite possible that some official tokens have the same or similar symbol names. For example, there are three ERC-20 tokens, HEDG (with only a symbol but no token name)\footnote{0xF1290473E210b2108A85237fbCd7b6eb42Cc654F}, Hedgie (HDG)\footnote{0x452B2bc7c94515720b36d304CE33909a8323F3e3} and Hedge (HDG)\footnote{0xfFe8196bc259E8dEDc544d935786Aa4709eC3E64}, that have similar names and symbols. Despite this, they are all independent and official tokens.
   An official token usually has its own official site, and we can collect detailed information about the token from Google. Thus, if a token has active transactions and we can collect its legal information online, we will regard it as a false positive and thus remove it from our dataset.
\end{itemize}

\begin{figure} [t]
\centering
  \includegraphics[width=0.99\textwidth]{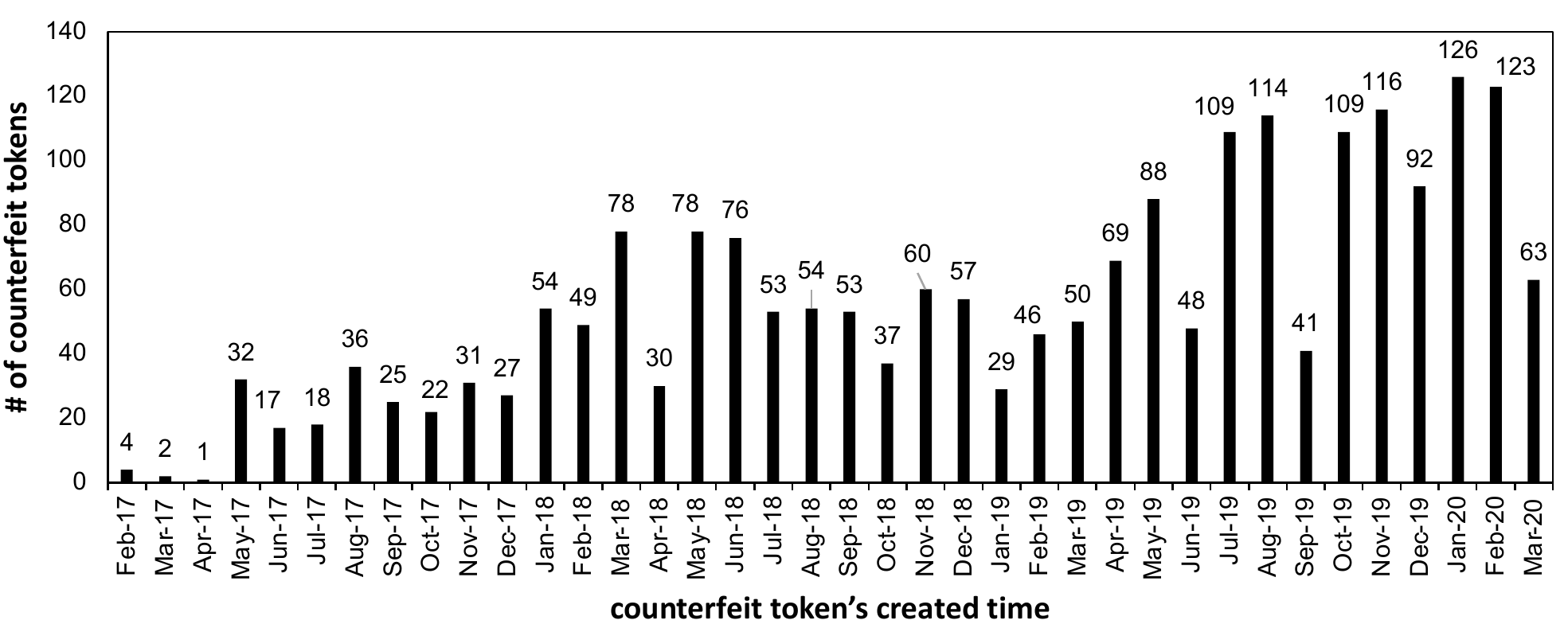}
\caption{The creation time of counterfeit tokens (till March 18th, 2020).}
\label{fig:fake token creation time} 
\end{figure}

\begin{figure} [t]
\centering
  \includegraphics[width=0.99\textwidth]{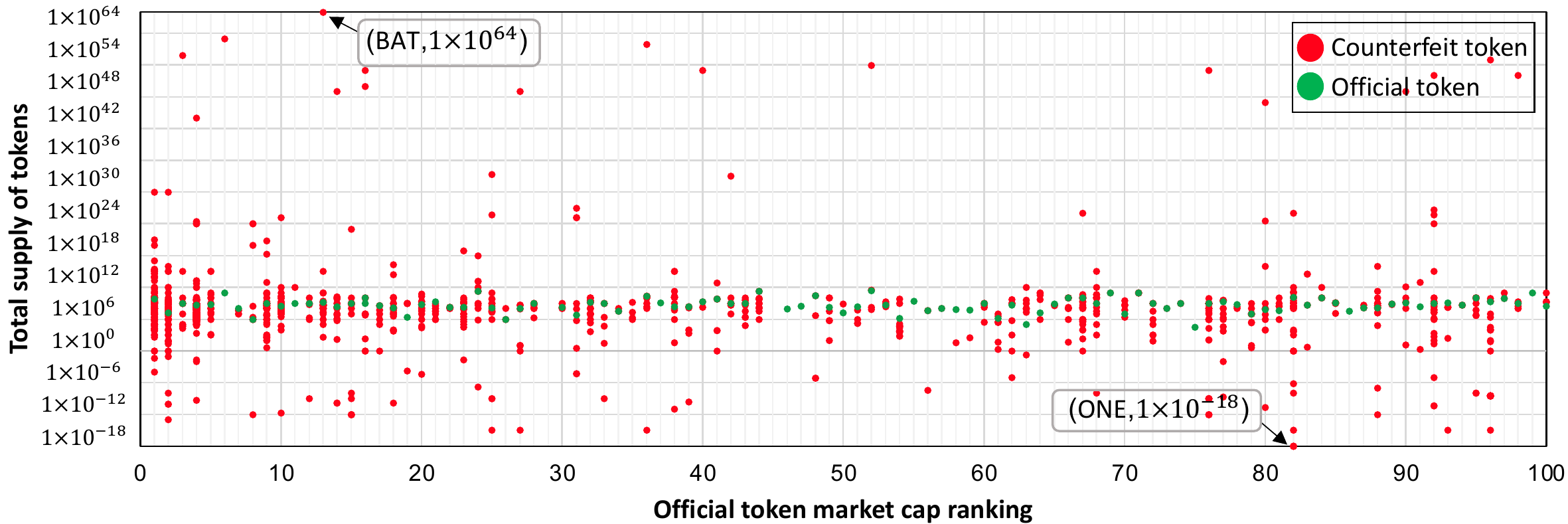}
\caption{The total supply of counterfeit tokens. Each vertical line represents the statistics of a type of token (100 lines in total). The dots shown on each vertical line indicate the total supply of an official token (in green) and its counterfeit ones (in red)}.
\label{fig:fake token creation total supply} 
\end{figure}

\subsection{Overall Results}

Using the above method, we have identified 2,117 counterfeit tokens, targeting 94 of the 100 popular tokens (94\%), as shown in Table~\ref{tab:targettoken}.
Unsurprisingly, HuobiToken (HT), Tether USD (USDT), and BNB (BNB) are the most popular targets. For example, there are 545 counterfeit tokens targeting HuobiToken (HT), with 12,883 transactions in total. HuobiToken (HT) is released by a famous cryptocurrency exchange with the same name Huobi, which is easier for unsuspecting users to believe its authenticity.
We observe that, \textit{in general, the counterfeit tokens are more likely to target popular tokens with high market capitalization rank}. However, not all high ranking tokens are their prior targets. For example, we only observe one counterfeit token that targets Crypto.com Coin (CRO), which ranks 6 by the time of our study.

Figure~\ref{fig:fake token creation time} shows the creation time of counterfeit tokens on a monthly basis.
The first counterfeit token was created at February 12nd, 2017. After that, counterfeit tokens have become increasingly prevalent, especially after July 2019. For example, in Jan 2020, 126 counterfeit tokens were created. 
\textit{This result suggests that, the counterfeit tokens are prevalent in the cryptocurrency ecosystem.}
There are 56,762 transactions related to these counterfeit tokens in total, involved 56,057 unique Ethereum addresses.
Figure~\ref{fig:fake token creation total supply} shows the comparison of \textit{total supply} between official tokens and their counterfeit tokens. Total supply refers to the number of coins or tokens in existence right now and are either in circulation or locked somehow~\cite{totalsupply}, which is defined by the token creator in the corresponding token contract. 
In Figure~\ref{fig:fake token creation total supply}, the dots on each line represent the total supply of a corresponding token (in green) and its counterfeit ones (in red).
For each official token, its total supply is usually between $1e+6$ and $1e+11$, but obviously, the total supply of some counterfeit tokens are very high. 
For example, the total supply of $100$ counterfeit tokens exceed $1e+12$. 
The highest one is BAT (BAT) \footnote{0x031f053a6b4e49A4A64450B2ca8A0b0ef6335134} which targets the official BAT (BAT), with a total supply of $1e+64$, while the total supply of official BAT (BAT) is only $1.5e+9$.

\subsection{Lexical Characteristics}
We next analyze the lexical characteristics of these 2,117 identified counterfeit tokens. As mentioned in Section~\ref{subsec:definition}, we consider two types of counterfeit tokens based on their naming strategies. 

\textbf{{Type-1.}}
As shown in Table ~\ref{tab:lexical characteristics}, almost 80\% of counterfeit tokens (1,674) have identical token names or symbols with official tokens. 
Among them, 23.6\% of counterfeit tokens (499) have exactly the same token names and symbols with official tokens, while 10.1\% of counterfeit tokens (214) have only the same token names with official tokens (but different symbols). Further, 45.4\% of counterfeit tokens (961) have the same symbols with official tokens (but different token names).

\textbf{{Type-2.}} Approximately 77\% of counterfeit tokens have adopted combo-squatting strategies in either their token names or their symbols. We characterize them into two categories. First, a number of counterfeit tokens combine the official identifier names with special characters like spaces, parentheses, underscores, or insert some abbreviations. For example, for the symbols of HuobiToken (HT), we have identified a number of confusingly similar symbols, including HT Coin, HT\_huobi, and Token HT, etc.
Second, the counterfeit tokens take advantage of different string encoding methods to mislead users, such as UTF-8, ASCII, and GBK, etc.

\begin{figure*}[t!]\centering 
\subfigure[Token name wordcloud]{ 
\begin{minipage}{6.5cm}\centering
\includegraphics[width = 0.8\textwidth]{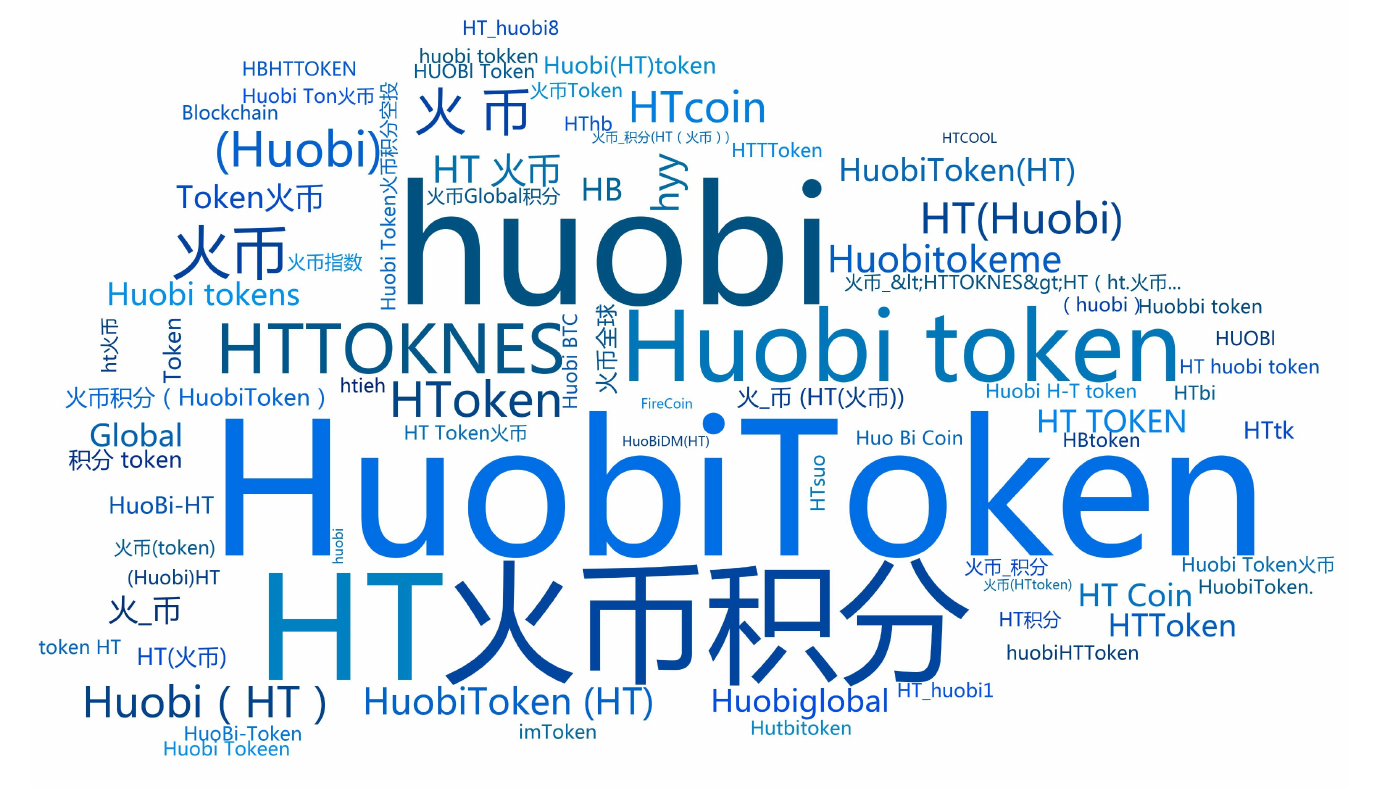} 
\end{minipage}}
\subfigure[Symbol wordcloud]{ 
\begin{minipage}{6.5cm}
\centering 
\includegraphics[width = 0.8\textwidth]{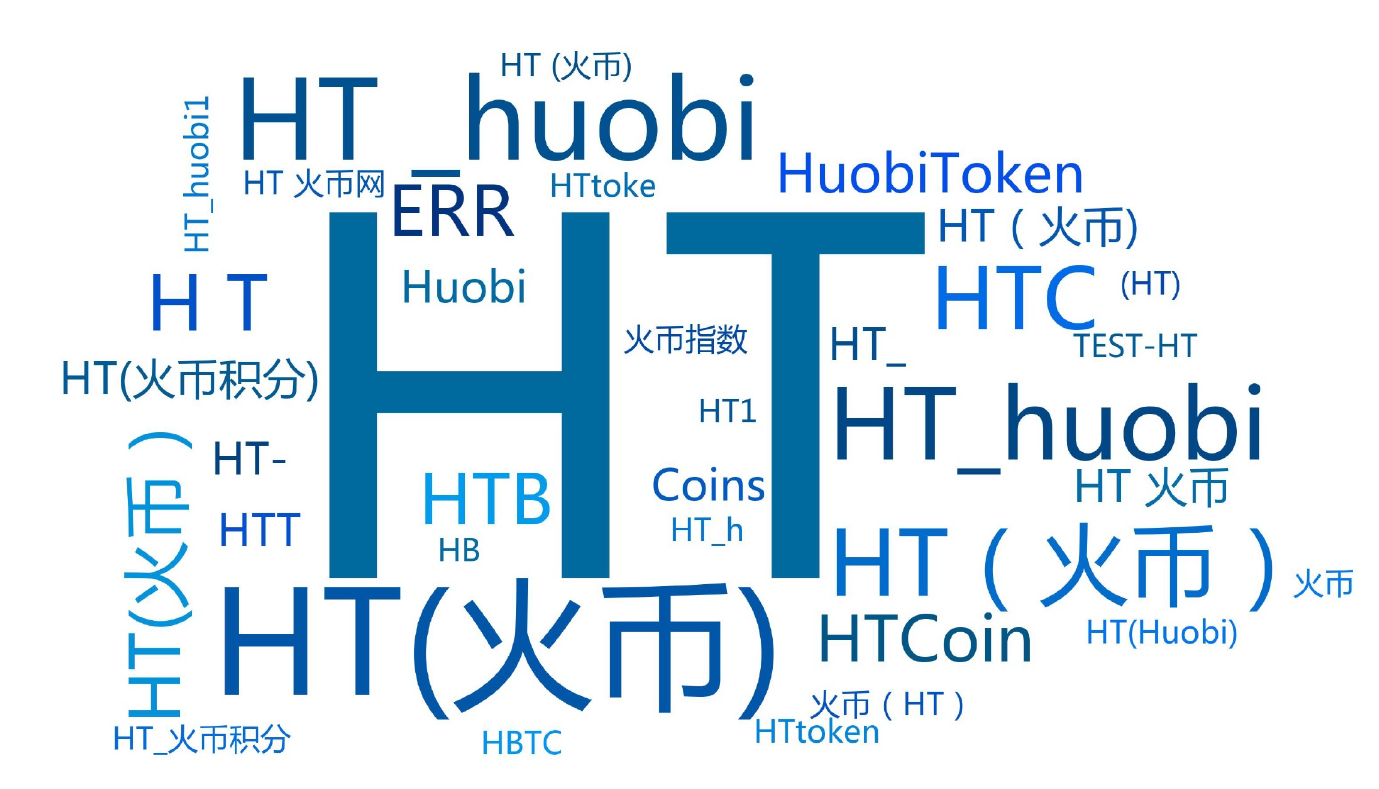} 
\end{minipage}
}
\caption{An example of the word cloud extracted from the 545 counterfeit tokens of HuobiToken (HT).}
\label{fig:fake token example}
\end{figure*}

Figure~\ref{fig:fake token example} shows an example of the word cloud extracted from the counterfeit tokens of HuobiToken (HT).
Note that the token names and symbols are shown separately. Besides the identical names with the official token, they always adopt a number of variants as aforementioned.

\begin{table}[t]
\caption{The lexical characteristics of counterfeit tokens.}
\label{tab:lexical characteristics}
\begin{tabular}{@{}ccc@{}}
\toprule
           & \# Combo-squatting (\%)   & \# Identical (\%)   \\ \midrule
Symbol     & 214 (10.1\%)     & 961 (45.4\%)    \\
Token name & 961 (45.4\%)     & 214 (10.1\%)   \\
Both       & 443 (20.9\%)     & 499 (23.6\%)  \\
Sum        & 1,618 (76.4\%)   & 1,674 (79.1\%) \\ \hline
All        & \multicolumn{2}{c}{2,117} \\ \bottomrule
\end{tabular}
\end{table}

\subsection{Popularity Analysis}

We next inspect the scale of these counterfeit currencies, looking at both \textit{the number of transactions} and \textit{the active period} of the tokens. 

\subsubsection{The number of transactions.}
\label{sec:transactionnumbers}
\begin{figure*}[htbp]\centering 
\subfigure[The number of counterfeit tokens' transactions.]{
\begin{minipage}{6.5cm}\centering
\includegraphics[width = 1\textwidth]{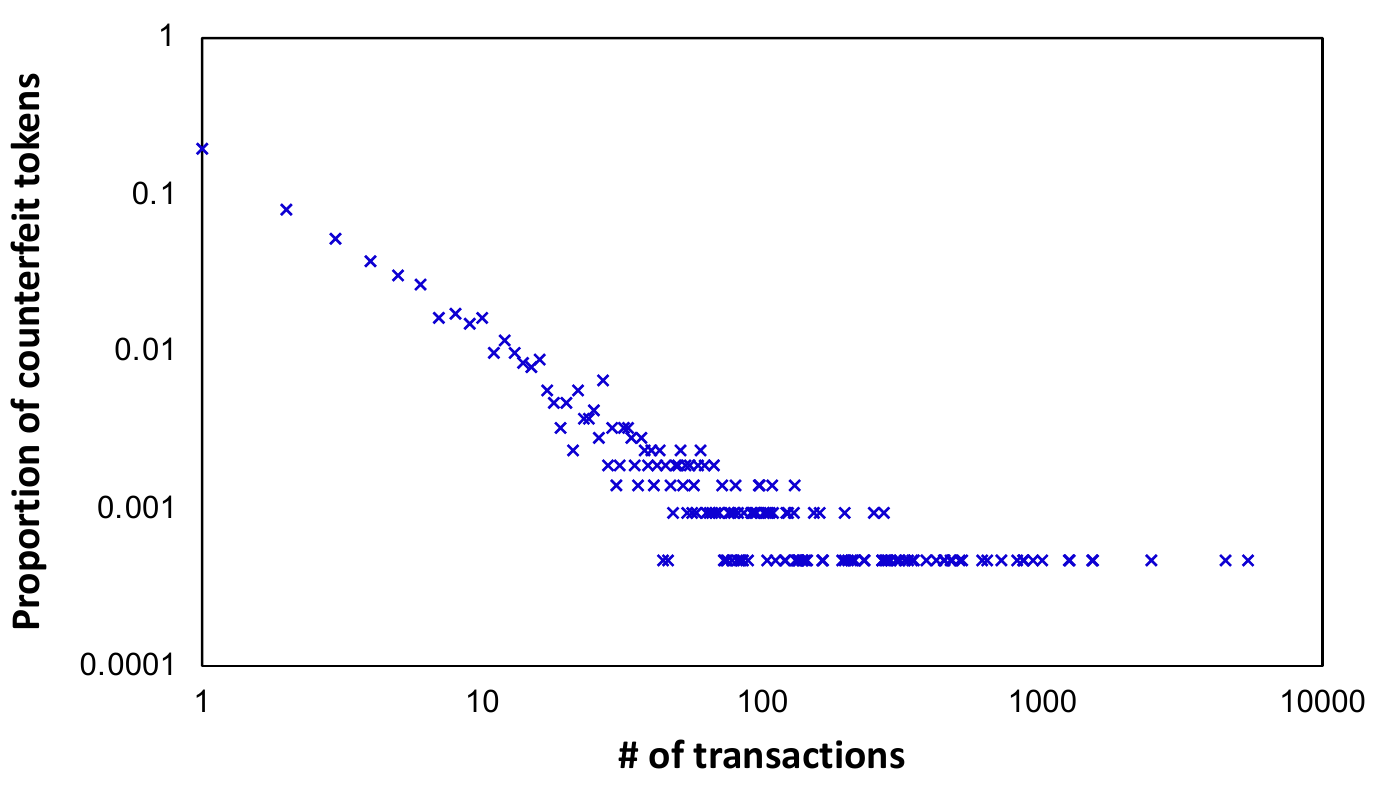} 
\end{minipage}}
\subfigure[Distribution of counterfeit tokens' active period.]{ 
\begin{minipage}{6.5cm}
\centering 
\includegraphics[width=1\textwidth]{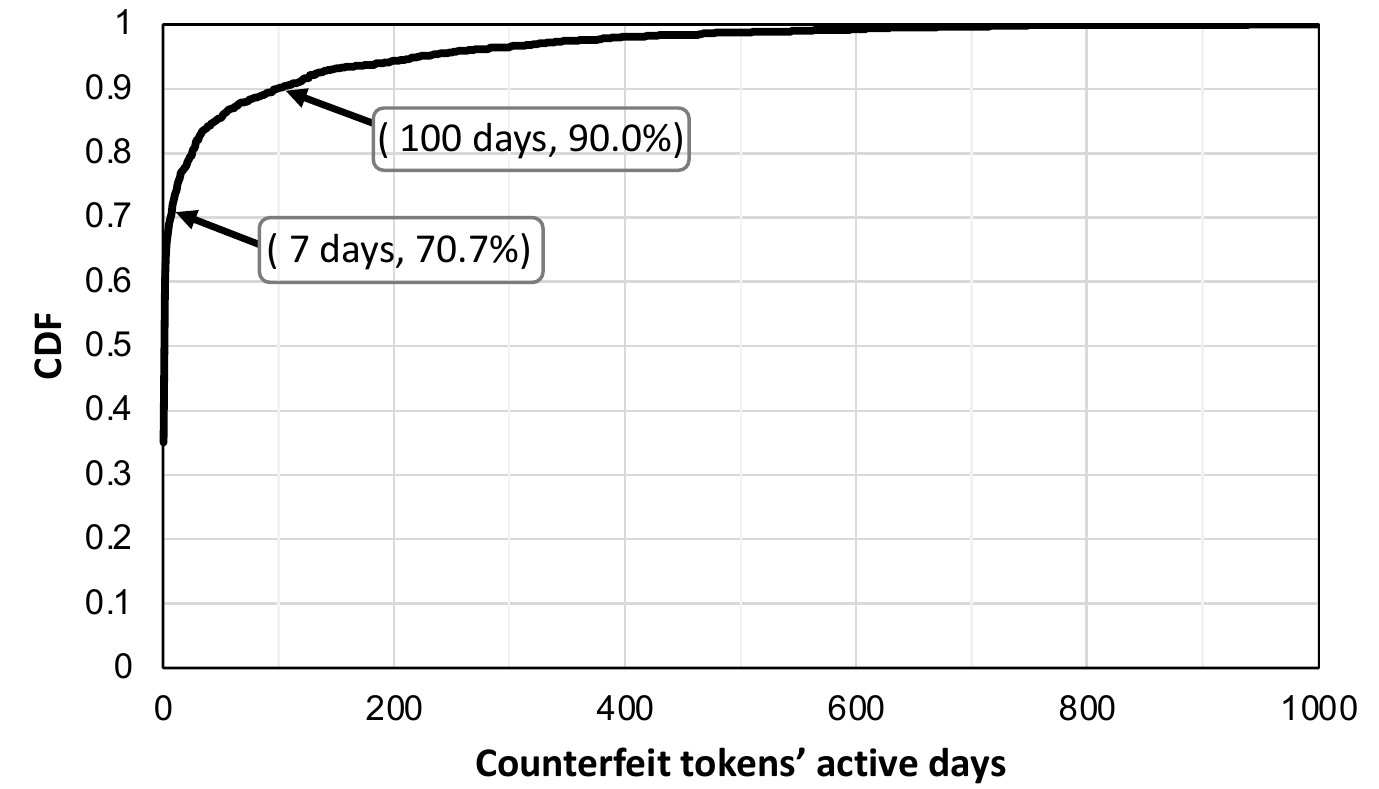} 
\end{minipage}
}
\caption{Distribution of counterfeit tokens' transactions and active period.}
\label{fig:fake token transaction and peroid}
\end{figure*}

The number of transactions can reflect the popularity of counterfeit tokens to some extent. 
Figure~\ref{fig:fake token transaction and peroid} (a) presents the number of transactions per token. We see that it varies greatly across the counterfeit tokens.
A large portion of the counterfeit tokens are less popular, i.e., 579 (27.35\%) counterfeit tokens have never been transferred, and over 90\% of counterfeit tokens have been transferred no more than 45 times. 
In contrast, some counterfeit tokens are very active with thousands of transactions. To be specific, 7 counterfeit tokens have over 1,000 transactions. 
Table~\ref{tab:top 5 fake token} shows the top-5 counterfeit tokens with the most transactions. 
The most heavily used counterfeit token, brc (brc), is an imitation token of Baer Chain (BRC), has over 5,000 transactions in total.
The counterfeit token Tether USD (USDT), also has over 4,500 transactions.
It is interesting to further investigate the activities related to these ``popular'' counterfeit tokens, which we will explore in Section~\ref{sec:scam}.

\begin{table}[]
\caption{The top-5 counterfeit tokens with the most number of transactions.}
\label{tab:top 5 fake token}
\resizebox{0.95\linewidth}{!}{
\begin{tabular}{@{}cccc@{}}
\toprule
Counterfeit Token &Target Token & Address  & Transactions \\ \midrule
brc (brc)&Baer Chain (BRC) & 0xB64555C4fEb7Dbe8584cb3b10D8993d1B3572f7e & 5,429               \\ 
Tether USD (USDT)&Tether USD (USDT)& 0x68771C9d7F6A498743Aa167967627b198B97e9E2 & 4,511               \\ 
USDT (Tether USD)&Tether USD (USDT)     & 0x3936eE7369e9c278A78a44dE7b272dEB97bc6253 & 2,446               \\ 
USDT (USDT)&Tether USD (USDT)& 0xfdF6b11baA0A17d39a378f0c97bC93dE8303f338 & 1,512               \\ 
POLYMATH (POLYM)&Polymath (POLY)    & 0x9e518098BB49354bc4372d48D3474d8C1F2eddF8& 1,503              \\ \bottomrule
\end{tabular}
}
\end{table}

\subsubsection{Active Period.}
We further analyze the active period of the counterfeit tokens. As the cost of creating a counterfeit token is quite low in Ethereum, we wonder \textit{whether the counterfeit tokens are only utilized for a short period}. 
Figure~\ref{fig:fake token transaction and peroid} (b) presents the distribution of the active time for all the counterfeit tokens. 
Over 70\% of the counterfeit tokens (1,497) are active for less than 7 days. However, to our surprise, some tokens remain active for a long time. Particularly, 9.97\% of the counterfeit tokens (211) are active over 100 days. 
For example, the counterfeit token 0x9900E9~\footnote{0x9900E95AE292e264B517f1979eB30C6c4D6458ab}, remains active for over 940 days, which is a BeraCoin (BRC) token that targets Baer Chain (BRC).

\subsection{Counterfeit Token Creator Analysis}

We next analyze the creators of the counterfeit tokens. Overall, 1,210 creators created 2,117 tokens. Over 95\% (2,016) of the tokens are created by external addresses (EOA, i.e., by humans, see Section~\ref{sec:background}), while the remaining tokens (101) are created automatically (COA, i.e., by smart contracts).

\begin{figure*}[h]\centering

\subfigure[Visualization of the Counterfeit Token Creator Graph.]{ 
\begin{minipage}{6.5cm}\centering
\includegraphics[width = 1\textwidth]{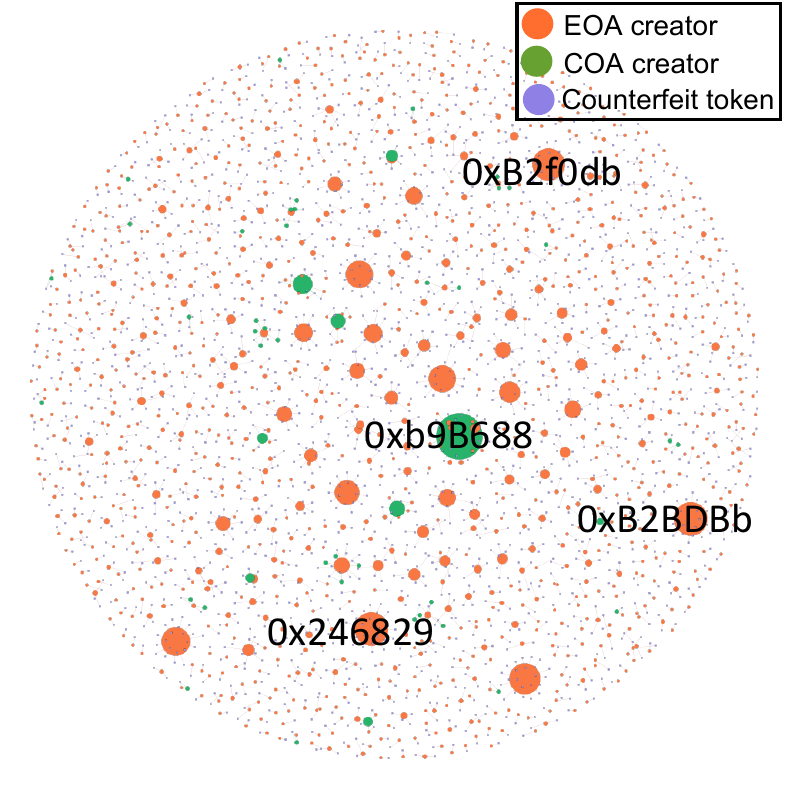} 
\end{minipage}}
\subfigure[The Overlap of Token Creators.]{ 
\begin{minipage}{6.5cm}
\centering
\includegraphics[width = 1\textwidth]{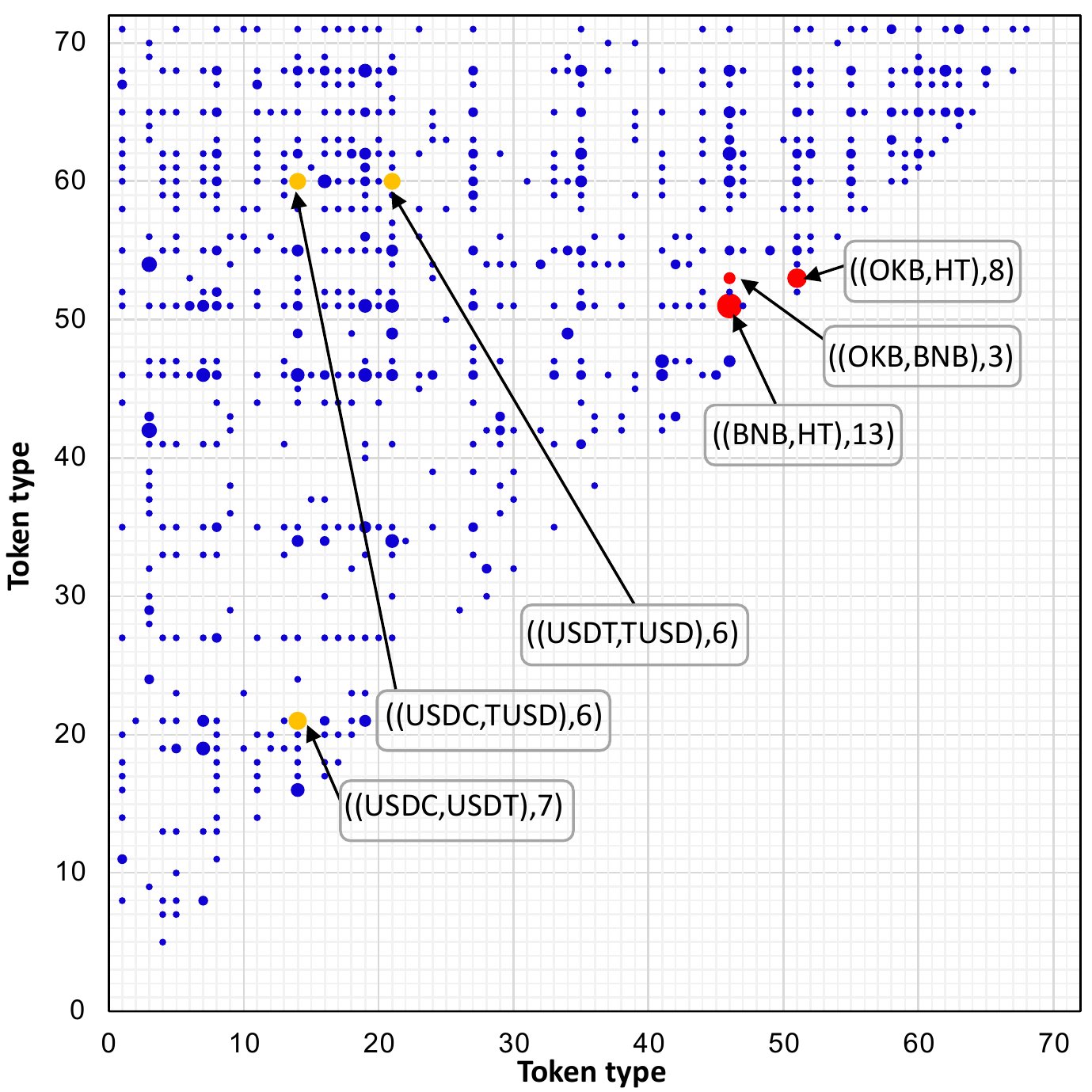} 
\end{minipage}
}
\caption{Visualization of Counterfeit Token Creators.}
\label{fig:fake token creator}
\end{figure*}

\textbf{Token Creator Graph.}
To further investigate the token creator relationship, we introduce the token creator graph (TCG), as shown in Figure~\ref{fig:fake token creator} (a).
In the TCG, each node denotes an address on Ethereum, i.e., the orange one represents the EOA creator address, the green one represents the COA creator address and the purple one is the counterfeit token address.
The TCG is a directed graph, with each edge represents the creation relationship, i.e., from EOA to counterfeit token, or from COA to counterfeit token. Note that, the COA need to be called by an EOA account to create a token, thus each COA has a connection with an EOA. The size of the creators' node denotes the number of counterfeit tokens they created.

\begin{table}[t]
\caption{The Top 5 Counterfeit Token Creators Ranked by the Number of Created Counterfeit Tokens.}
\label{tab:top 5 creators}
\resizebox{0.9\linewidth}{!}{
\begin{tabular}{c|c|c}
\toprule
Address & \begin{tabular}[c]{@{}c@{}}\# CTokens\end{tabular} & \begin{tabular}[c]{@{}c@{}}\# CToken Types\end{tabular} \\ \midrule
0x2468293D8059bc578CF312F09Ed78D6CE1005dCb & 25 & 23 \\ 
0xB2BDBb9Cc6583D10C3c043DE3AC6bE07A074dd16 & 25 & 1 \\ 
0xB2f0dbb7FF8f1A451fF9486756EA53Ff2F654633 & 24 & 1 \\ 
0x1f7A74F06359e08Cc82a5c60cFa21D277f5f4181 & 23 & 21 \\ 
0xb9B6885D0Af9914d432871DcBeB20DAa8282A763 & 23 & 5 \\ \bottomrule
\end{tabular}
}
\end{table}

We observe that almost 30\% of creators (362) have released more than one counterfeit token. 
This shows that \textit{certain actors may specialize in the creation of such tokens}.
Table~\ref{tab:top 5 creators} presents the top-5 addresses that create the most counterfeit tokens. 
The most aggressive addresses are  0xB2BDBb\footnote{0xB2BDBb9Cc6583D10C3c043DE3AC6bE07A074dd16} and 0x246829\footnote{0x2468293D8059bc578CF312F09Ed78D6CE1005dCb}, which have both created 25 counterfeit tokens.
By further analyzing the creation of counterfeit token, we find that \textit{most creators (1,107) exclusively counterfeit the same currency}.
However, 103 creators counterfeit multiple currencies. For example, the most aggressive one, 0x246829\footnote{0x2468293D8059bc578CF312F09Ed78D6CE1005dCb}, has counterfeited 23 official tokens (with 25 counterfeit tokens released in total). 

Thus, we analyze the co-occurrence of identical token creators across different types of counterfeit tokens. As shown in Figure~\ref{fig:fake token creator} (b), there are 71 types of targeted tokens\footnote{For 23 official tokens, their counterfeit tokens do not share creators with other tokens, thus we eliminate them in the figure to save space. The orders of tokens in Figure~\ref{fig:fake token creator} (b) are: BRC, OECAN, QKC, SNT, DGD, ZB, BTM, LRC, REN, HEDG, INB, TRAC, AION, USDC, HOT, PAX, NPXS, LINK, OMG, THETA, USDT, AGI, GNO, LEO, LEND, XIN, ELF, OGN, NEXO, MATIC, KNC, TRB, VEN, DAI, IOST, CEL, ANKR, IOTX, ONE, STORJ, ZIL, SEELE, UBT, CRO, ENJ, BNB, ICX, FET, SAI, UTK, HT, BNT, OKB, FTM, MKR, DATA, KCS, POWR, ENG, TUSD, GNT, BAT, MANA, NOAH, REP, SXP, POLY, ZRX, WBTC, CHZ, WAX.}. All of the 71 official tokens have counterfeit tokens whose creators have generated other kinds of counterfeit tokens. The size of the circle represents the number of creators counterfeited both of the corresponding tokens. We mark two sets of tokens with the most identical creators. They are HuobiToken (HT), BNB (BNB) and OKB (OKB) in red, and Tether USD (USDT), True USD (TUSD) and USD Coin (USDC) in orange. Among them, HuobiToken (HT) and BNB (BNB) share the most counterfeit token creators (13). The following is HuobiToken (HT) and OKB (OKB), with 8 counterfeit token creators in common. 
We speculate that scammers tend to create counterfeit tokens with tokens with the same characteristics. First, they are all top-50 tokens with high popularity. For HuobiToken (HT), BNB (BNB) and OKB (OKB), they are all released by well known cryptocurrency exchanges that share identical names: HuobiToken (HT) is released by Huobi, BNB (BNB) is released by Binance (with same Chineses pronunciation and spelling), and OKB (OKB) is released by OKB. For Tether USD (USDT), True USD (TUSD) and USD Coin (USDC), they are all stablecoins (stable-value cryptocurrency) that mirror the price of the U.S. dollar.
\textit{This result suggests that malicious actors tend to counterfeit more than one type of official token, likely due to the low cost of creating tokens on Ethereum}.

\begin{figure*}[h]\centering 
\subfigure[Visualization of Counterfeit Token Holder]{ 
\begin{minipage}{6.5cm}\centering
\includegraphics[width = 0.7\textwidth]{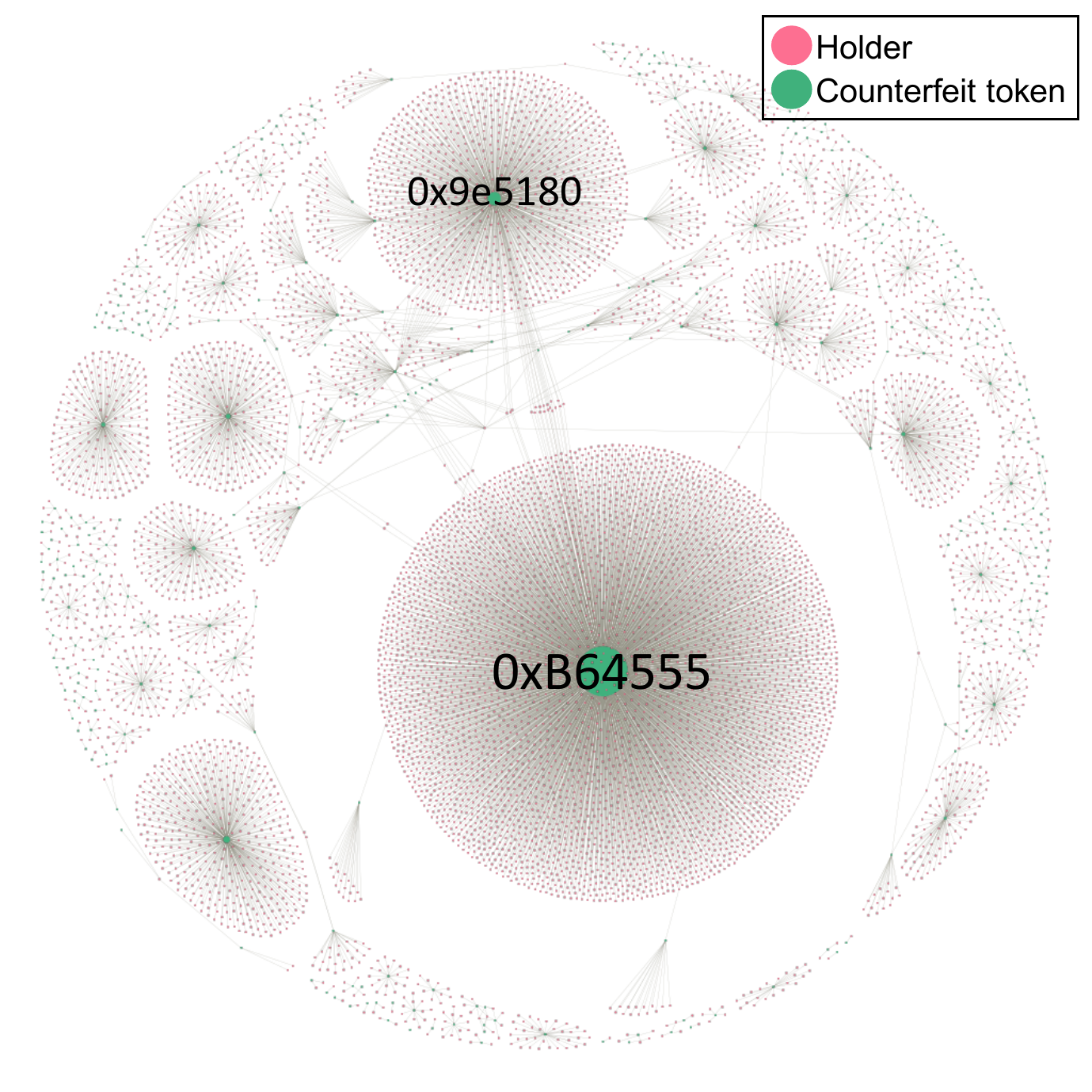} 
\end{minipage}}
\subfigure[Distribution of counterfeit tokens' holders]{ 
\begin{minipage}{6.5cm}
\centering 
\includegraphics[width = 0.95\textwidth]{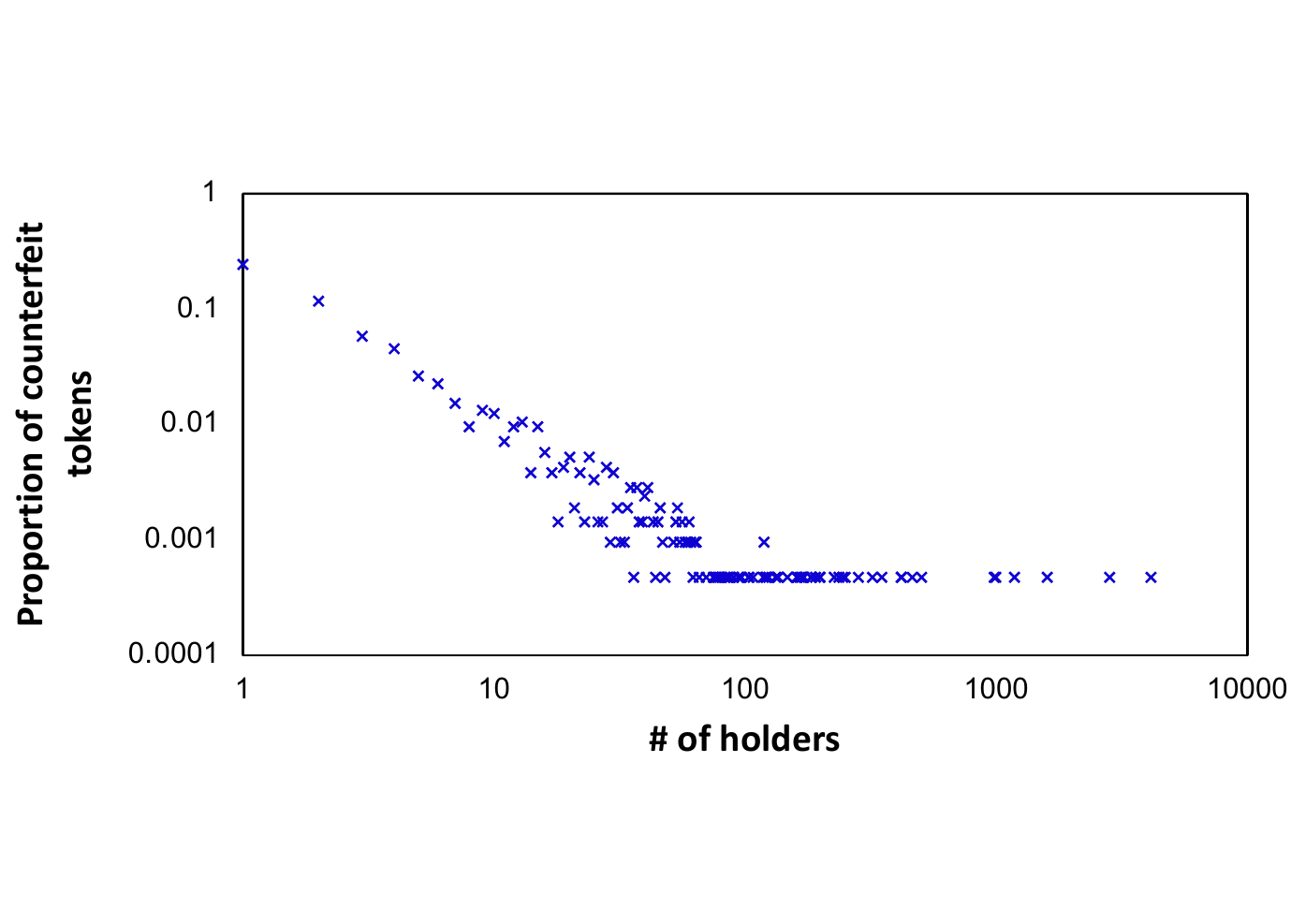} 
\end{minipage}}
  \vspace{-0.1in}
\caption{Visualization of Counterfeit Token's Holders.}
  \vspace{-0.1in}
\label{fig:fake token holder}
\end{figure*}

\subsection{Counterfeit Token Holder Analysis}

We next analyze the characteristics of token holders, the cornerstone of the ecosystem. Overall, there are 28,861 unique holders in total. 
As the number of holders is much higher than the number of creators, we use a sampled (20\% of all the holders) counterfeit token holder graph (THG) for clear illustration, as shown in Figure~\ref{fig:fake token holder} (a).
Note that our sampling method is based on the proportion of holders for each counterfeit token.
Figure~\ref{fig:fake token holder} (b) presents the distribution of holders of the counterfeit tokens.
Over half of the tokens (64.58\%) have no more than two holders. Over 90\% of the tokens have no more than 20 holders. It further suggests that, most tokens are only used for a short time, thus have limited holders and transactions to prevent victims from reporting or causing suspicion from regulators. The token with the most holders (4,147) is 0xB64555\footnote{0xB64555C4fEb7Dbe8584cb3b10D8993d1B3572f7e}. 
We further observe that, although over 96\% of holders only possess a single kind of counterfeit token, roughly 1,189 holders have multiple kinds of counterfeit tokens. Among them, 506 holders have multiple kinds of counterfeit tokens that target more than one official currency.
Table~\ref{tab: top 5 holders} shows the top-5 holders ranked by the number of different counterfeit tokens. 
The largest holder address,  0x8d12A1\footnote{0x8d12A197cB00D4747a1fe03395095ce2A5CC6819}, holds a remarkable 85 different counterfeit tokens that target 47 types of official currencies.

\begin{table}[h]
\caption{The top-5 token holders ranked by the number of different counterfeit tokens.}
\label{tab: top 5 holders}
\resizebox{0.7\linewidth}{!}{
\begin{tabular}{@{}ccc@{}}
\toprule
Address of counterfeit token holder & \# CTokens & \begin{tabular}[c]{@{}c@{}}\# CToken Types\end{tabular} \\ \midrule
0x8d12A197cB00D4747a1fe03395095ce2A5CC6819 & 85 & 47 \\
0xB2BDBb9Cc6583D10C3c043DE3AC6bE07A074dd16 & 24 & 1  \\
0x09f91Ce790dbb36ddA4EC507A4754b7a4e2b0e55 & 21 & 21 \\
0xa4fd7ACa0A39e1c70d464D6380e293761d64FA63 & 20 & 19 \\
0xb9A8436700cbaBbf855d03a00513502515C49B83 & 18 & 4  \\ \bottomrule
\end{tabular}
}
\end{table}

\begin{figure} [t]
\centering
  \includegraphics[width=0.9\textwidth]{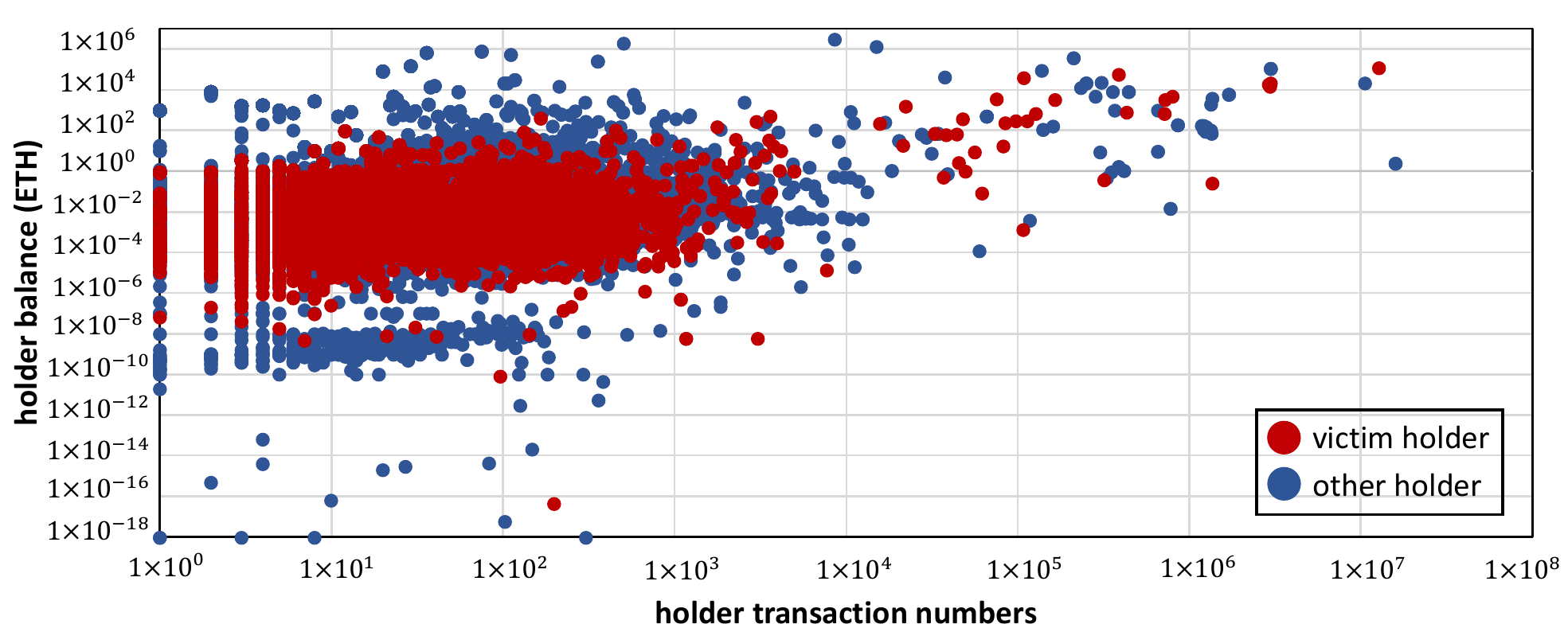}
  \vspace{-0.1in}
\caption{The distribution of the balance and the number of transactions of counterfeit token holders.}
  \vspace{-0.1in}
\label{fig:holder} 
\end{figure}

We further investigate the type of users that are generally affected by counterfeit cryptocurrencies. Note that it is non-trivial to determine the experience of counterfeit token holders, yet we can gain insights from the number of transactions per holder. In general, the more transactions related to the holder address, the more experienced the holder is. Figure~\ref{fig:holder} shows the distribution of the number of the transactions and the balance of counterfeit token holders. Note that, some holders are verified as victims (see Section~\ref{sec:scam}), which are labelled in red.
Almost 50\% of holders (13,812) have fewer than 5 transactions, and nearly 95\% of holders (27,134) have a balance under 0.2 ETH. 
Thus, the observation is inline with our expectation that most of the holders (victims) are novices without much experiences.

\begin{framed}
\noindent \textbf{Answer to RQ1:} 
\textit{
Counterfeit tokens are indeed prevalent in the cryptocurrency ecosystem. We have identified 2,117 counterfeit tokens targeting 94 out of the 100 tokens we studied. Although most of the tokens have very few transactions, some of them are quite popular, with thousands of transactions and holders. Malicious actors tend to target more than one type of official token, mainly due to the cost of creating a counterfeit token is quite low in Ethereum.
}
\end{framed}

\section{Fraudulent Behaviors of Counterfeit Cryptocurrencies}
\label{sec:scam}

Our prior exploration suggests the prevalence of counterfeit cryptocurrencies in Ethereum.
However, we are still unaware of \textit{the usage of counterfeit tokens}. Although the token names and symbols can be counterfeit in Ethereum, the token addresses cannot. The token address is the unique identifier to represent a token. As long as users input the official address during the transaction, the counterfeit cryptocurrency scams will not succeed.
Thus, in this section, we seek to explore the scams and social engineering attacks related to counterfeit tokens.

\subsection{Types of Fraudulent Behaviors}

To understand the characteristics of scams related to counterfeit tokens, we first resort to existing scam reports to identify the types of scam activities related to counterfeit tokens. To be specific, we first resort to the following scam repositories, and use keyword searching (e.g., ``fake'', ``token'' and ``counterfeit'') to identify related scams for manual verification.

\begin{itemize}
    \item[1)] \textbf{CoinHunter.} CoinHunter~\cite{coinhunter} is a crowd-sourcing platform that collects cryptocurrency related scams reported from users. We have implemented a crawler to get all the scam reports from the CoinHunter. After keyword filtering and manual inspection, 52 reports related to counterfeit tokens on Ethereum are identified.
    
    \item[2)] \textbf{Blockchain Forums.} 
    The imTokenFans~\cite{imtoken} is a forum operated by the official team of imToken, a popular cryptocurrency wallet. BitcoinTalk~\cite{bitcointalk} is an online forum devoted to the discussion of Bitcoin and other cryptocurrencies.
    Both of them host the ``Scam Accusations'' board for users to report scams.
    Thus, we have crawled all the related posts on the ``Scam Accusation'' board based on aforementioned keywords and further perform a manual verification. At last, we have identified 18 reports on counterfeit token scams.
    
    \item[3)] \textbf{Other scam reports from search engines.} We also resort to search engines to identify scams related to the identified counterfeit tokens. We have implemented an automated crawler, to feed the 2,117 counterfeit token addresses to Google and get all the search results. After eliminating unrelated ones (e.g., blockchain address searching services), we manually verify whether there are user complaints, and media reports on these counterfeit token addresses. This helps us identify 10 more scam reports.
\end{itemize}

At last, by analyzing the collected scam reports, we have collected 204 counterfeit token addresses, all of which have been included in our 2,117 counterfeit token dataset.
The identified scams can be classified into two types: \textit{airdrop scam} (22) and \textit{arbitrage scam} (182). 
\textit{Note that the identified scams are only used as the ground truth for our following study. We will further propose approaches to investigate how many counterfeit tokens have such fraudulent behaviors in the Section~\ref{subsec:airdrop} and Section~\ref{subsec:arbitrage}, as the scams are typically under-reported by users in the wild.}

\subsubsection{Airdrop Scam}
\begin{figure*}[t]\centering 
\subfigure[The information the fraudster posted]{ 
\begin{minipage}{6cm}\centering
\includegraphics[width = 0.96\textwidth]{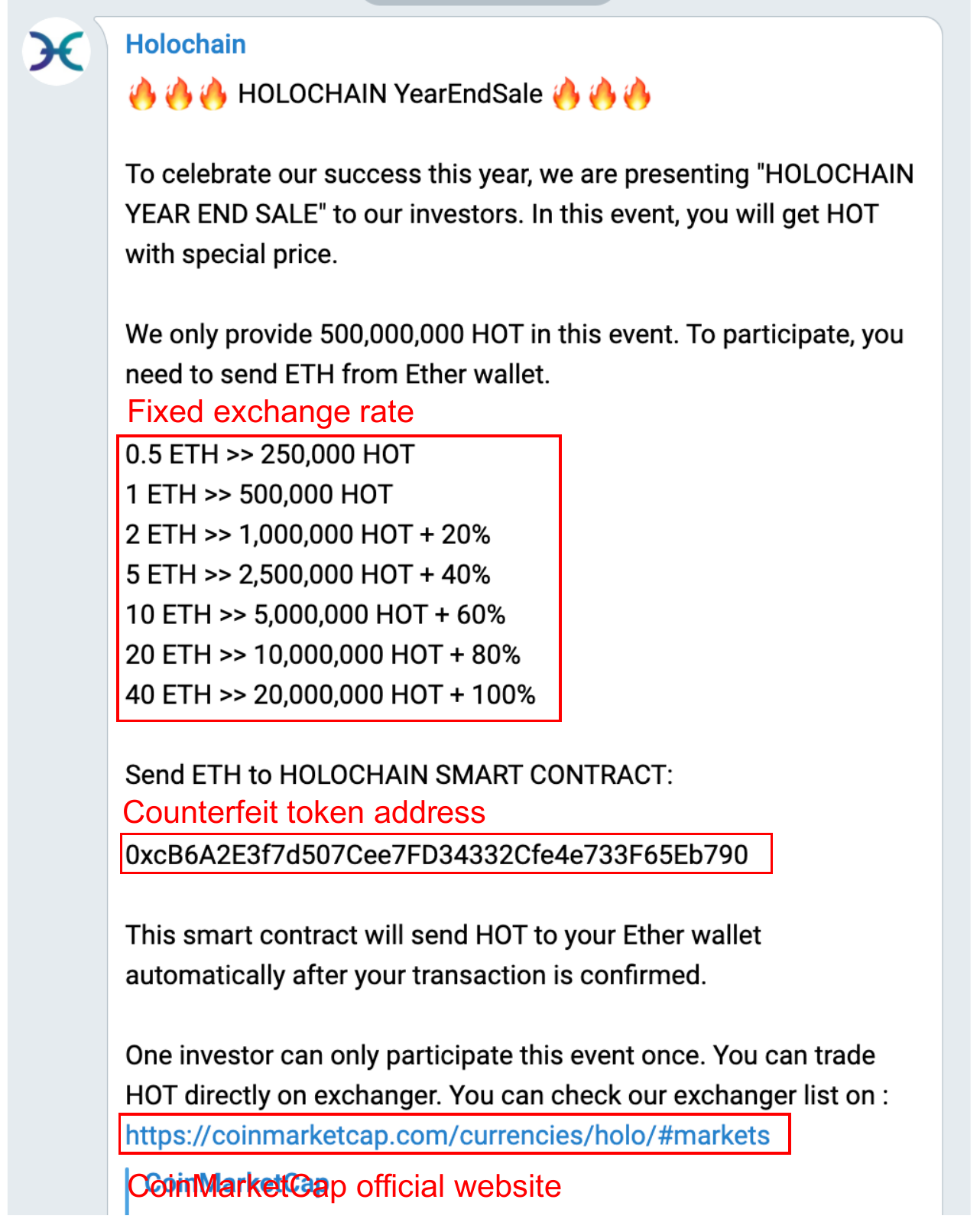} 
\end{minipage}}
\subfigure[The screenshot the fraudster provided]{ 
\begin{minipage}{6cm}
\centering 
\includegraphics[width = 1.1\textwidth]{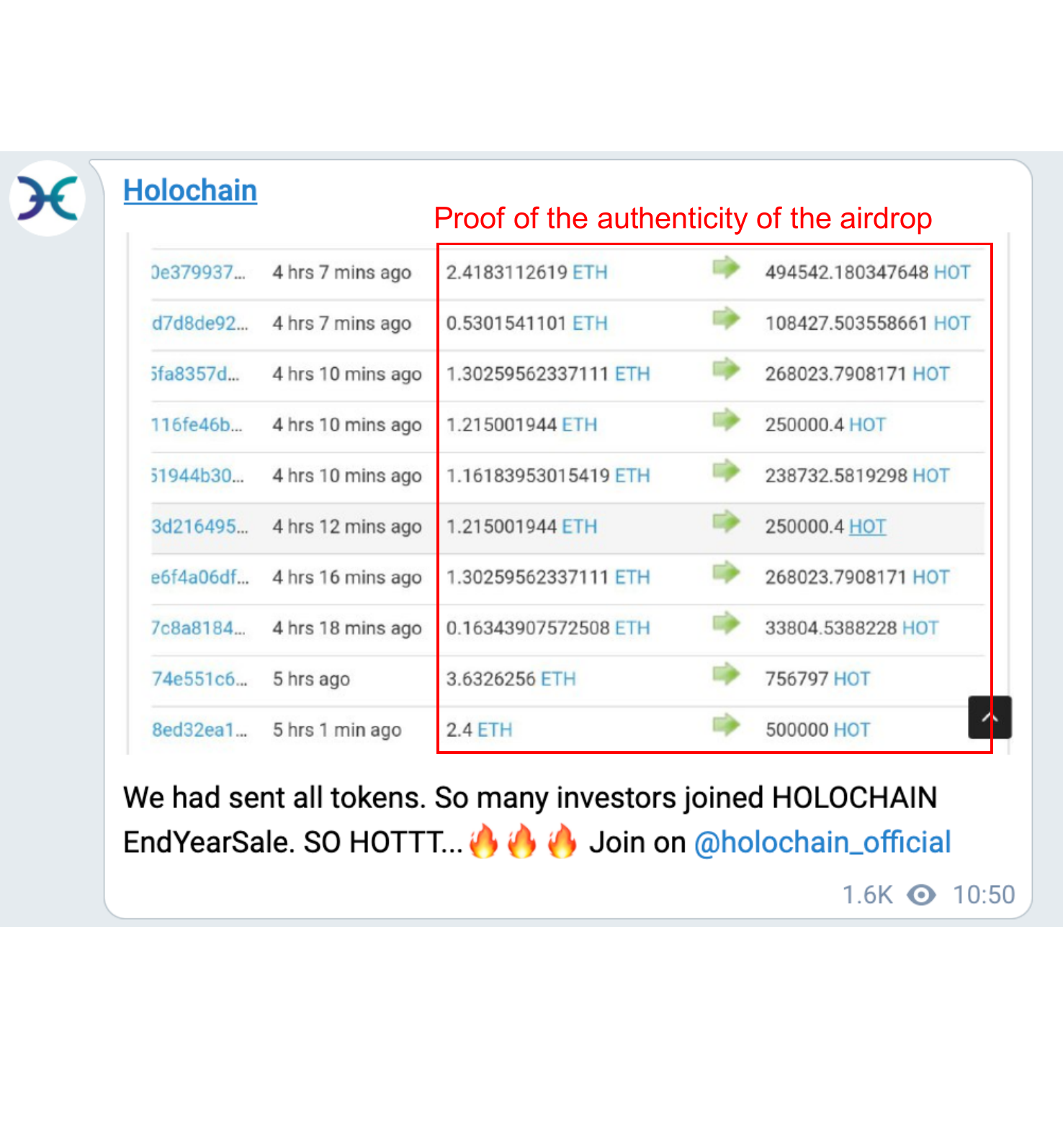} 
\end{minipage}
}
\caption{An example of a fake HOLOTOKEN (HOT) airdrop scam.}
\label{fig:airdrop information}
\end{figure*}

An airdrop is the distribution of a cryptocurrency token, \textit{usually for free}, to numerous user wallet addresses~\cite{wiki-airdrop}. 
Airdrops are primarily implemented as a way of gaining attention and new followers.
As airdrops are quite popular for well-known tokens, counterfeit tokens are also taking advantage of this opportunity to perform airdrop scams.
In general, the attackers promise that, after sending a certain amount of ETH to the (counterfeit) token address, the victim will get (imitated) official tokens according to a \textit{fixed exchange rate} (far more than the actual value). After victims send the ETH, they likely only receive counterfeit tokens that have no value at all.
Figure~\ref{fig:airdrop information} shows an airdrop scam of the counterfeit token HOLOTOKEN (HOT), with the same icon and account name of the official token.
To deceive unsuspecting users, they usually embed the official link of the imitated token in the airdrop information (see Figure~\ref{fig:airdrop information} (a)).
Furthermore, the scammer will also post screenshots of the relevant tokens returned to the user to enhance the credibility (but the tokens they returned are fake) (see Figure~\ref{fig:airdrop information} (b)).

\subsubsection{Arbitrage Scam}
Arbitrage is an investment method that capitalizes on imbalances in prices between markets, i.e., buy at a low price and sell at a slightly higher price~\cite{ArbitrageExplanation}.
From the perspective of amateur investors, the arbitrage does not require too much professional knowledge, which is relatively safer compared with other investment methods. Thus, cryptocurrency arbitrage is popular among many investors.
However, our exploration suggests that the arbitrage can be abused by attackers, i.e., \textit{the arbitrage can be combined with counterfeit tokens to carry out well-designed scams}.
In our collected scams, the scammers usually use fake Telegram groups (that imitate the official token) as advertising channels (see Section~\ref{sec:advertising}), providing scam addresses for victims to send ETH. After victims send ETH to the specified scam address, as promised, they should get official tokens (far more than the actual value) in a few minutes. However, victims usually received counterfeit tokens of no worth.

Next, we will analyze in detail the workflow of these two scams, devise approaches to detect such scams, and measure the prevalence of the scams in our 2,117 counterfeit token dataset.

\subsection{Analyzing and Detecting Airdrop Scam}
\label{subsec:airdrop}

\subsubsection{The Workflow of Airdrop Scams.}
\begin{figure*}[t]\centering 
\subfigure[The overall process of the airdrop scam]{ 
\begin{minipage}{6cm}\centering
\includegraphics[width = 1\textwidth]{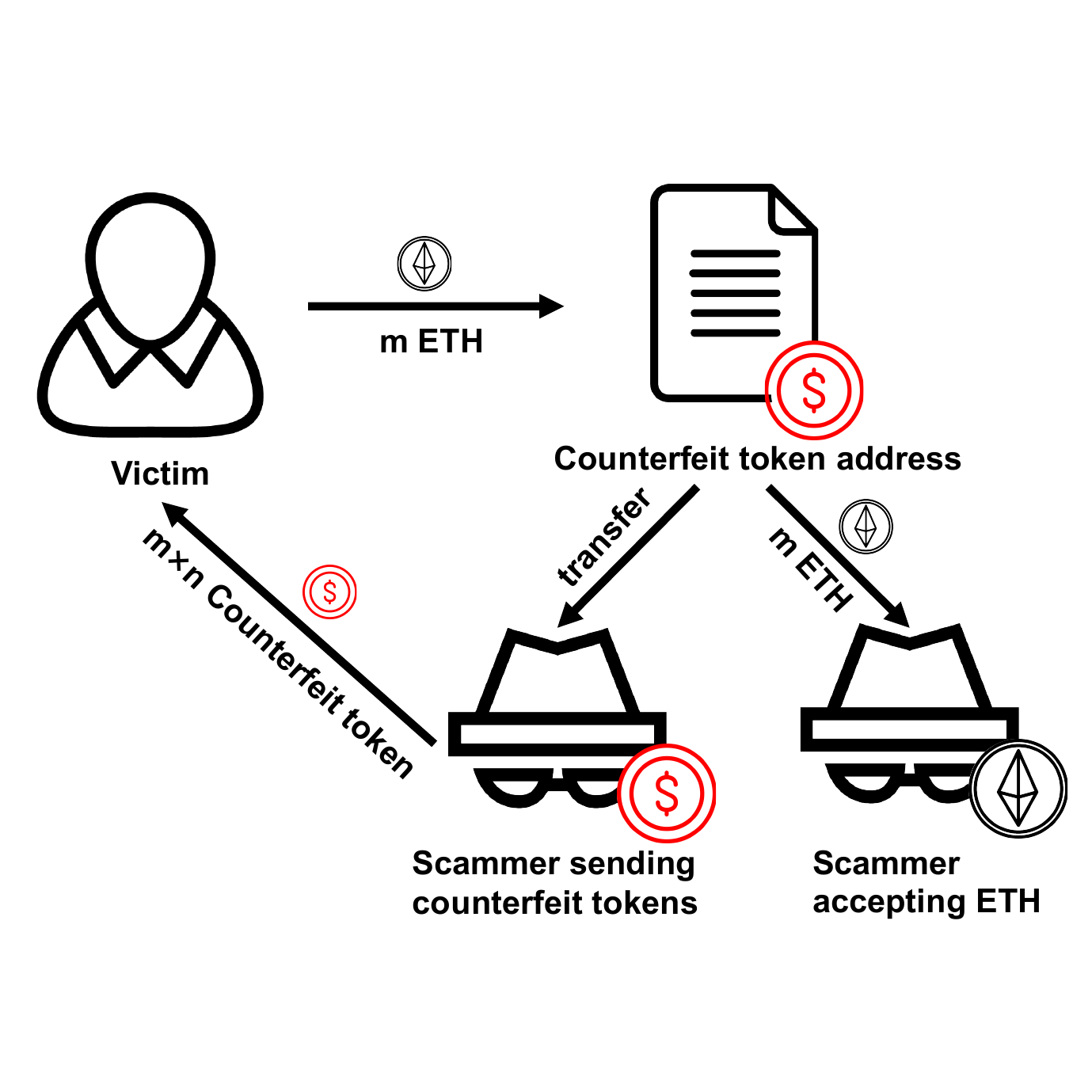} 
\end{minipage}}
\subfigure[The transaction graph of an airdrop scam]{ 
\begin{minipage}{6cm}
\centering
\includegraphics[width = 1\textwidth]{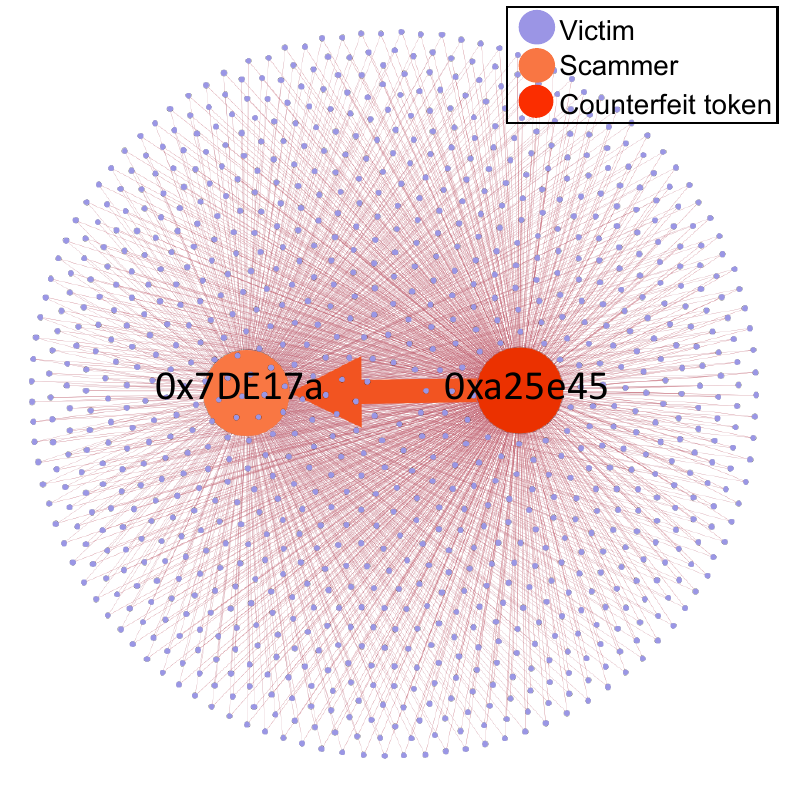} 
\end{minipage}
}
\caption{Visualization of the Airdrop Scam.}
\label{fig:airdrop}
\end{figure*}

Figure~\ref{fig:airdrop} (a) shows the workflow of airdrop scams, which consists of four main roles: \textit{the victim}, \textit{the counterfeit token contract}, \textit{the scam address that accepts ETH for money laundering}, and \textit{the scam address that sends counterfeit tokens}. Note that in the airdrop scam, all these steps are fulfilled in one transaction.
When \textit{the victim} sends \textit{m} ETH to \textit{the counterfeit token contract},
the contract transfers all the received ETH to the scam address that accepts ETH, and then calls the transfer function to return \textit{m×n} counterfeit tokens to the victim, distributed by the scam address that sends counterfeit tokens. The exchange rate of ETH to counterfeit token is fixed (the \textit{n}) in Figure~\ref{fig:airdrop} (a). 
Note that the scam address that accepts ETH and the scam address that sends counterfeit tokens can be either same or different addresses.

\subsubsection{Detecting Airdrop Scams.}
Based on the characteristics of airdrop scams, we seek to investigate whether the 2,117 counterfeit tokens we identified have such behaviors. To be specific, for a transaction involving airdrop scams, \textcircled{1} \textit{the victim address must send some ETH to the counterfeit token contract first}; and \textcircled{2} the counterfeit token distributor must transfer some fixed percentage of counterfeit tokens to the victim. Furthermore, \textcircled{3} the counterfeit token contract should transfer the received ETH to a scam address. Thus, we have implemented a script to analyze all the transactions related to counterfeit tokens. 
A counterfeit token whose transactions meet the aforementioned behaviors will be considered as part of an airdrop scam.

\subsubsection{Results and Observations.}
This method can achieve 100\% accuracy in our collected ground truth dataset, i.e., all the 22 collected airdrop scams are correctly detected. As airdrop scams have explicit patterns, i.e., the exchange rate of ETH to counterfeit token is fixed, our detection has no false positives.
We further apply this method to all 2,117 counterfeit tokens we identified, i.e., analyzing their related transactions to detect airdrop scams.
As a result, 87 counterfeit tokens have shown the behaviors of airdrop scams, targeting 44 official tokens.
Overall, 2,037 victims were deceived in airdrop scams, and the attackers have received 970.8 ETH in total (\$$226,817.71$)\footnote{As the price of Ether fluctuates everyday, we estimate the profit using the closing price of Ether on July 16th.}.
Table~\ref{tab:airdrop_example} shows the top-10 counterfeit tokens with airdrop scams ranked by the number of ETH they received. Note that, for each of them, the scam address used to receive ETH and the scam address used to distribute counterfeit tokens are identical (see Column ``scam address'' in Table~\ref{tab:airdrop_example}).
The counterfeit token 0xd7bb68\footnote{0xd7bb68B0cE5893983e5a2511b87E083609eB6fF9} gains the most profit in the airdrop scam, reaching 180.46 ETH (\$$42,163.14$). 
The counterfeit token Polymath (POLY), whose address is 0xa25e457~\footnote{0xa25e457ae666A67C9C906d1052fAE39710Af63CC7}, has the most number of victims (133). 
Figure~\ref{fig:airdrop} (b) shows its transaction graph. The red point denotes the Polymath (POLY) counterfeit token address, the orange point denotes the scam address used to receive ETH and distribute counterfeit tokens, and the purple points denote the victims. The size of points represent the amount of ETH involved, and the thickness of the edge represents the amount of ETH or counterfeit tokens transferred. 
Besides, it is obvious that the exchange rate of ETH to counterfeit tokens is stable (ranging from 330 to 1,125,000, see the Column ``Token/ETH'' in Table~\ref{tab:airdrop_example}).

\begin{table}[t]
\caption{The top-10 counterfeit tokens involved in airdrop scams ranked by the number of ETH received.}
\resizebox{\linewidth}{!}{
\begin{tabular}{|ccccccc|}
\hline
Token &
  Contract &
  Token/ETH &
  Victims &
  Scam Address &
  \begin{tabular}[c]{@{}c@{}}ETH\\ Received\end{tabular} &
  \begin{tabular}[c]{@{}c@{}}Est.\\ Value (\$)\end{tabular} \\ \hline
 QuarkChain Token (QKC)& 0xd7bb68B0cE5893983e5a2511b87E083609eB6fF9 & 32,000 & 16 & 0x7F83284C0cE0906F58eFF0FC433967e50c4D607E & 180.46 & 42,163.14\\ \hline
 Matic Token (MATIC)   & 0x4256117a02aC880335f8bFbEed63f92eC0001A5a &53,992.  &10 & 0xCC9CD9d0CA055616093c52741F19a01Fb8fD709c & 107.76 &  25,176.51\\ \hline
HOLOTOKEN (HOT)  & 0xcB6A2E3f7d507Cee7FD34332Cfe4e733F65Eb790 & 1,125,000 & 24  & 0xe4F349f54e8490b7aE9EFB090Be7EAA41b08D965   & 75.0 & 17,523.00 \\ \hline
Enjin Coin (ENJ)  & 0x7e534b4192daeaa6559c08c7147364b00a7ce697 & 5,000    & 109 & 0xe937910a2A748296eE884fC2beb1A041789DA50F   & 70.2 & 16,401.53 \\ \hline
 Matic Token (MATIC) & 0xd7E6460a5ff2e51BD1583808d1C19d521CF43DeF & 91,255 & 20 & 0x3A3870d906f6B8EF1127E78103B038939B58b63e  & 62.62 & 14,630.84\\ \hline
 Matic Token (MATIC) & 0x41993b3F7979dccD8A15e1448CCACCEbC6873803&91255 & 13 & 0xBb733A9c379cA22829cb689C9eafC04c0ea8A51f     & 58.19 & 13,596.12\\ \hline
 Matic Token (MATIC) & 0x4b67BE266D81BF5CEE60FC12b3bded87fCDaC783 & 47,015 & 49 & 0xB663b4BeA1197e58D69F27B04f8ee2B7531e9668  & 57.16 & 13,355.20\\ \hline
Holotoken (HOT)  & 0x5566E98b8Bce420E943C7366ab0F15ba3D181F10 & 1,000,000 & 27  & 0x23705f9b708eE6363Dcc93Cc3d8056098a3c5787   & 36.2 & 8,457.77 \\ \hline
Polymath (POLY) & 0xa25e457ae666A67C9C906d1052fAE39710Af63CC & 10,000 & 133 & 0x7DE17a49f857dA070D4a6c6d272Def7B6e603015 & 34.74 & 8,117.57 \\ \hline
OmiseGo (OMG)  & 0xFdDe0a92bfEf2d95695d32400A0A8F55e530dC67 & 330     & 33  & 0x0531B689856db9B745f804Bd0506f443244E4981 & 33.9 &  7,920.40\\ \hline
Kyber Network (KNC)  & 0x29ad674d180D33f0391b7b043d4880918b66b72e & 1,215   & 36  & 0x97FECFC329f5213DABA1a173519F89026c33A3eD & 30.0  &  7,009.20\\ \hline
\end{tabular}
}
\label{tab:airdrop_example}
\end{table}

\subsection{Analyzing and Detecting Arbitrage Scams}
\label{subsec:arbitrage}

\subsubsection{The Workflow of Arbitrage Scams}
\begin{figure}[t]
\centering
  \includegraphics[width=0.8\textwidth]{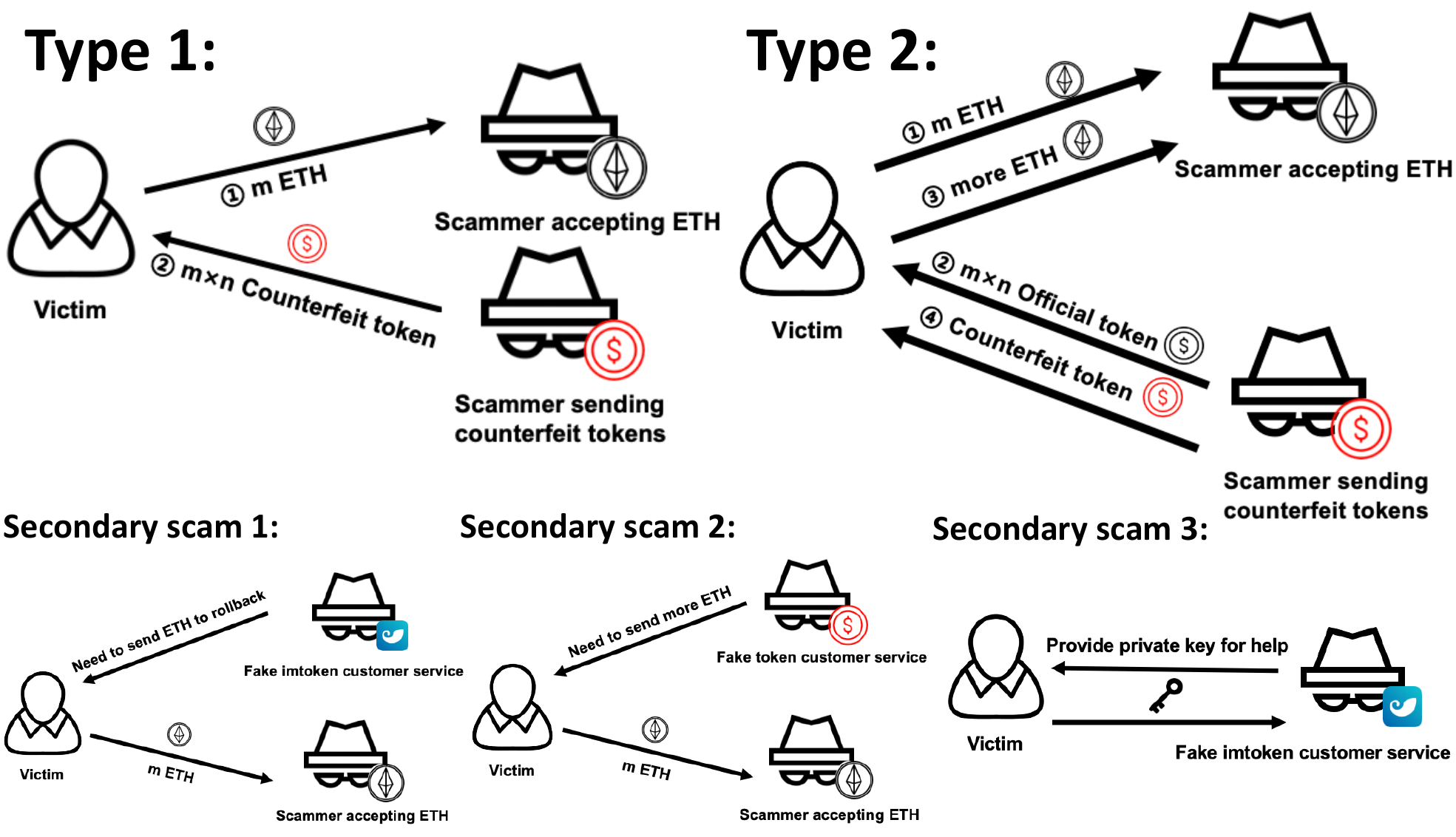}
\caption{The workflow of arbitrage scams.}
\label{fig:arbitrage characteristics} 
\end{figure}

Figure~\ref{fig:arbitrage characteristics} shows the workflow of arbitrage scams, which consists of four main roles: \textit{the victim}, \textit{the scam address that accepts ETH}, \textit{the scam address that sends counterfeit tokens}, and \textit{the scammer posing as the official customer service} (e.g., a fake Telegram account). 
Based on our collected scams, we have summarized two types of arbitrage scams: 1) scammers return counterfeit tokens directly after receiving the ETH; 2) scammers return official tokens at first, but then return counterfeit tokens in the following transactions.

\textbf{Type 1:}  The victim sends ETH to the address that the scammer provides, and receives counterfeit tokens in a few minutes or does not receive tokens at all. 
Since counterfeit tokens are manual sent to the victim by the scammer, the scammer can easily conduct a \textit{secondary scam}. If the victim does not receive tokens at all, the fake imToken (cryptocurrency wallet) customer service will explain that the smart contract address cannot read the victim address successfully. The victim needs to send the same amount of ETH to the same scam address for rollback to get the ETH back or provides the private key for help (see secondary scam 1 and 3 in Figure~\ref{fig:arbitrage characteristics}). If the victim receives counterfeit tokens and identifies that they could not use counterfeit tokens to trade on the exchange, the counterfeit token customer service will explain that the amount of tokens is too small to use the exchange for withdrawal, and the victim needs to send more ETH to get tokens. 

\textbf{Type 2:} The victim sends ETH to the address that the scammer provides, and receives official tokens in a few minutes for the first time. The victim would then believe the arbitrage's authenticity and send more ETH to the scam address, but will only receive counterfeit tokens the second time.

\subsubsection{Detecting Arbitrage Scams.}
Unlike airdrop scam, we cannot identify the arbitrage scam directly through the counterfeit token contract, as they do not accept ETH directly. 
Thus, we devise an approach that consists of the following steps.
First, we regard every transfer of the counterfeit token as a \textit{candidate transfer from the scammer to the victim}. 
Thus, the receiver address in the transfer could be a potential victim address. 
Note that one major feature of the arbitrage scam is that, before victims receive counterfeit tokens, they should have already sent the ETH to the scam address a while ago (we found that almost all the victims received counterfeit tokens from the confirmed scam addresses within 2 hours after they sent ETH in our collected scams). 
Thus, we further identify the corresponding ETH transfer transaction of the potential victims. 
The ETH transfer transaction should be the most recent one just before the counterfeit token transfer. As a result, the receiver address in the ETH transfer should be the scam address published in Telegram or other platforms (see Section~\ref{sec:advertising}). In this way, we can accurately identify the arbitrage scams and label both the victim addresses and scam addresses.

\subsubsection{Results and Observations.}
The above method can achieve a recall of 97.8\% in our collected ground truth dataset, with 4 missed cases.
The reason is that we track the scam address through the transfer of counterfeit tokens, thus \textit{we cannot confirm the scam address that did not send the counterfeit token to the victim, i.e., the victim receives nothing at all in this scenario}. 
We further apply this method to all the 2,117 counterfeit tokens.
We have identified 7,617 transfer transactions related to 486 counterfeit tokens (targeting 10 official tokens) that are considered to be arbitrage scams.
We further identify 1,879 scam addresses that are used to accept ETH.
To measure the potential false positives of our arbitrage scam detection, we manually sampled 100 identified scam addresses to investigate whether some of them are scams. On the one hand, we check the identified scam addresses on Google to find whether they were involved in any scam posts (e.g., telegram groups). On the other hand, we manually analyzed their transactions to see whether they follow some specific patterns. For example, the real scam addresses usually have transactions with multiple victims. If most of their transactions follow the patterns we summarized (note that normal addresses would not have such behaviors), we believe they are definitely scams. Our manual verification does not flag any false positives.
The total financial losses is 73,300.9 ETH (\$$17,126,022.30$).
Table~\ref{tab:arbitrage tokens} shows the statistics of the counterfeit tokens that are involved in arbitrage scams. Obviously, HuobiToken (HT) has the largest scale of counterfeit token arbitrage scams, whose campaigns have received $60,824$ ETH (\$$14,210,919.40$). 

\begin{table}[t]
\caption{Counterfeit tokens that involved in arbitrage scams.}

\resizebox{0.85\linewidth}{!}{
\begin{tabular}{@{}lrrrrr@{}}
\toprule
\begin{tabular}[c]{@{}l@{}}Targeted  token\end{tabular} &
  \begin{tabular}[c]{@{}l@{}}\# CTokens\end{tabular} &
  \begin{tabular}[c]{@{}l@{}}\# transactions\end{tabular} &
  \begin{tabular}[c]{@{}l@{}}\# victims\end{tabular} &
  \begin{tabular}[c]{@{}l@{}}ETH received\end{tabular} & 
  \begin{tabular}[c]{@{}l@{}}USD received\end{tabular}\\ \midrule
HuobiToken (HT)   & 373 & 4,291 & 2,610 & 60,824.3 & 14,210,989.50 \\
USD Coin (USDC) & 17  & 809  & 578  & 4,559.1  & 1,065,188.12\\
OKB (OKB)  & 20  & 295  & 245  & 2,948.7  & 688,934.27\\
BNB (BNB)  & 12  & 258  & 156  & 2,265.7  & 529,358.15\\
TrueUSD (TUSD) & 4  & 162   & 95   & 1,428.7  & 333,801.47\\
Paxos Standard (PAX)  & 5  & 130   & 82   & 967.0   & 225,929.88\\
Tether USD (USDT) & 36 & 1,377  & 1,074 & 226.2   & 52,849.37\\
Bytom (BTM)  & 13  & 265  & 254   & 68.6    & 16,027.70\\
Zilliqa (ZIL)  & 3   & 18   & 15    & 12.0    & 2,803.68\\
IOSToken (IOST) & 3   & 12   & 10   & 0.6      & 140.18\\ \midrule
Sum  & 486 & 7,617 & 5,087 & 73,300.9 & 17,126,022.30\\ \bottomrule
\end{tabular}
}
\label{tab:arbitrage tokens}
\end{table}

\textbf{Secondary scam.}
We observe that over 29\% of victims (1473) have been deceived more than once. 
Note that the real proportion can be higher as the attacker sometimes asks the victim to transfer ETH to other addresses that we cannot track. 
The victim that was scammed the most is 0xE5D1Ef~\footnote{0xE5D1Ef3297896FAA8B118031edDDC7a372655932}, who has transferred ETH to the scam address 19 times, with a total amount of $2,799.1$ ETH (\$$635,981.72$).

\textbf{Type-1 and Type-2 scams.}
We are also interested in the victims who have ever received official tokens from confirmed scam addresses. 
4.6\% of victims have received official tokens from the scam address (\textit{Type-2 scam}). 
Then, 81\% of them send ETH to the scam addresses for the second time.
This result suggests that \textit{sending official tokens to the victim can greatly increase the probability for victims to send ETH to the scam address again}.
When sending ETH for the second time, over 90\% of victims send more ETH compared with the amount they sent at first time.
 
For example, the victim 0xf8a6aa~\footnote{0xf8a6aa3fcEE296DE9c388492c496aAa85EA66eee} sent 1 ETH (\$$223.64$) to 0xa6C678~\footnote{0xa6C678Ed8b54521Bf0DC46933aFb48005f543411} (the scam address) for the first time, and it received 55 official HuobiToken (HT). 
After that, the victim sent 115.18 ETH (\$$26,910.66$) to the scam address one hour later, but only received counterfeit tokens in the second time.

\begin{table}[]
\caption{Summary of the scams we identified.}
\resizebox{0.85\linewidth}{!}{
\begin{tabular}{c|c|c|c|c|c|c}
\hline
Type                                                                      & \# Transactions        & \multicolumn{2}{c|}{Scam Address}                                                        & \# Victims              & \# ETH                    & \# USD                        \\ \hline
\multirow{5}{*}{\begin{tabular}[c]{@{}c@{}}Airdrop \\ Scam\end{tabular}}  & \multirow{5}{*}{2,872}    & Counterfeit Token Address                                                        & 87    & \multirow{5}{*}{2,037}   & \multirow{5}{*}{970.8}    & \multirow{5}{*}{226,817.71}     \\ \cline{3-4}
                                                                          &                         & Counterfeit Token Creator                                                        & 71    &                        &                           &                               \\ \cline{3-4}
                                                                          &                         & ETH Received Address                                                             & 70    &                        &                           &                               \\ \cline{3-4}
                                                                          &                         & \begin{tabular}[c]{@{}c@{}}Counterfeit Token \\ Distributed Address\end{tabular} & 56    &                        &                           &                               \\ \cline{3-4}
                                                                          &                         & Sum                                                                              & 166   &                        &                           &                               \\ \hline
\multirow{5}{*}{\begin{tabular}[c]{@{}c@{}}Arbitrage\\ Scam\end{tabular}} & \multirow{5}{*}{7,617} & Counterfeit Token Address                                                        & 486   & \multirow{5}{*}{5,087} & \multirow{5}{*}{73,300.9} & \multirow{5}{*}{17,126,022.30} \\ \cline{3-4}
                                                                          &                         & Counterfeit Token Creator                                                        & 293   &                        &                           &                               \\ \cline{3-4}
                                                                          &                         & ETH Received Address                                                             & 1,879 &                        &                           &                               \\ \cline{3-4}
                                                                          &                         & \begin{tabular}[c]{@{}c@{}}Counterfeit Token \\ Distributed Address\end{tabular} & 869   &                        &                           &                               \\ \cline{3-4}
                                                                          &                         & Sum                                                                              & 2,904 &                        &                           &                               \\ \hline
Sum                                                                       & 10,489                  & \multicolumn{2}{c|}{3,053 (565 counterfeit tokens involved)}                             & 7,104                  & 74,271.7                 & 17,352,840.00                  \\ \hline
\end{tabular}
}
\label{tab:scam volume}
\end{table}

\subsection{Summary of the Scams}

\subsubsection{The Impact of the Scams.}
We summarize the overall impact of the scams, including the \textit{involved addresses}, the \textit{financial losses}, and \textit{the number of victims}. As shown in Table~\ref{tab:scam volume}, we see 565 counterfeit tokens are involved in an airdrop (87) or arbitrage scam (486). As we mentioned in Section ~\ref{sec:transactionnumbers}, over 27\% of counterfeit tokens (579) have never been transferred, and almost 75\% of counterfeit tokens (1,584) were transferred fewer than 9 times. We have manually analyzed many less popular counterfeit tokens and found it hard to justify their usage based on no or very few transactions. These less popular tokens may be used for testing or used in some scams, but no users were successfully cheated. Thus, we cannot infer their original intentions based on no or few transactions. We classify scam addresses into four categories based on their roles: \textit{counterfeit token contract}, \textit{counterfeit token creator}, \textit{the money laundering address that receives ETH}, and \textit{the counterfeit token distribution address}. 
From all aspects, the scale of arbitrage scam is much larger than that of airdrop. In general, the airdrop scam addresses are more likely to play more than one role. On average, each scam address has 1.71 roles. While for the arbitrage scam addresses, their roles are more specific (i.e., each scam address has 1.21 roles on average).
\textit{Overall, 7,104 victims have been successfully cheated, with an overall financial loss of 74,271.7 ETH (\$$17,352,840.00$)}, which is a lower bound estimate of the criminal profits. Interestingly, 20 victims were deceived by both of the scams. We further tagged all victims who held counterfeit tokens in Figure~\ref{fig:holder}. It is interesting to see that only 67.2\% of victims (4,775) still hold counterfeit tokens after they have been cheated. We infer that after victims receive counterfeit tokens, they usually try to transfer these tokens to other addresses to obtain ETH. We observe that most of the victims are novices to Ethereum, i.e., have limited transactions and balance. Among these victims, over 5\% of victims (2,518) have less than 20 transactions, and 92.8\% of victims (4,431) have balance less than 0.2 ETH.

\subsubsection{Money flow of scam addresses.}
\begin{figure} [t]
\centering
  \includegraphics[width=0.6\textwidth]{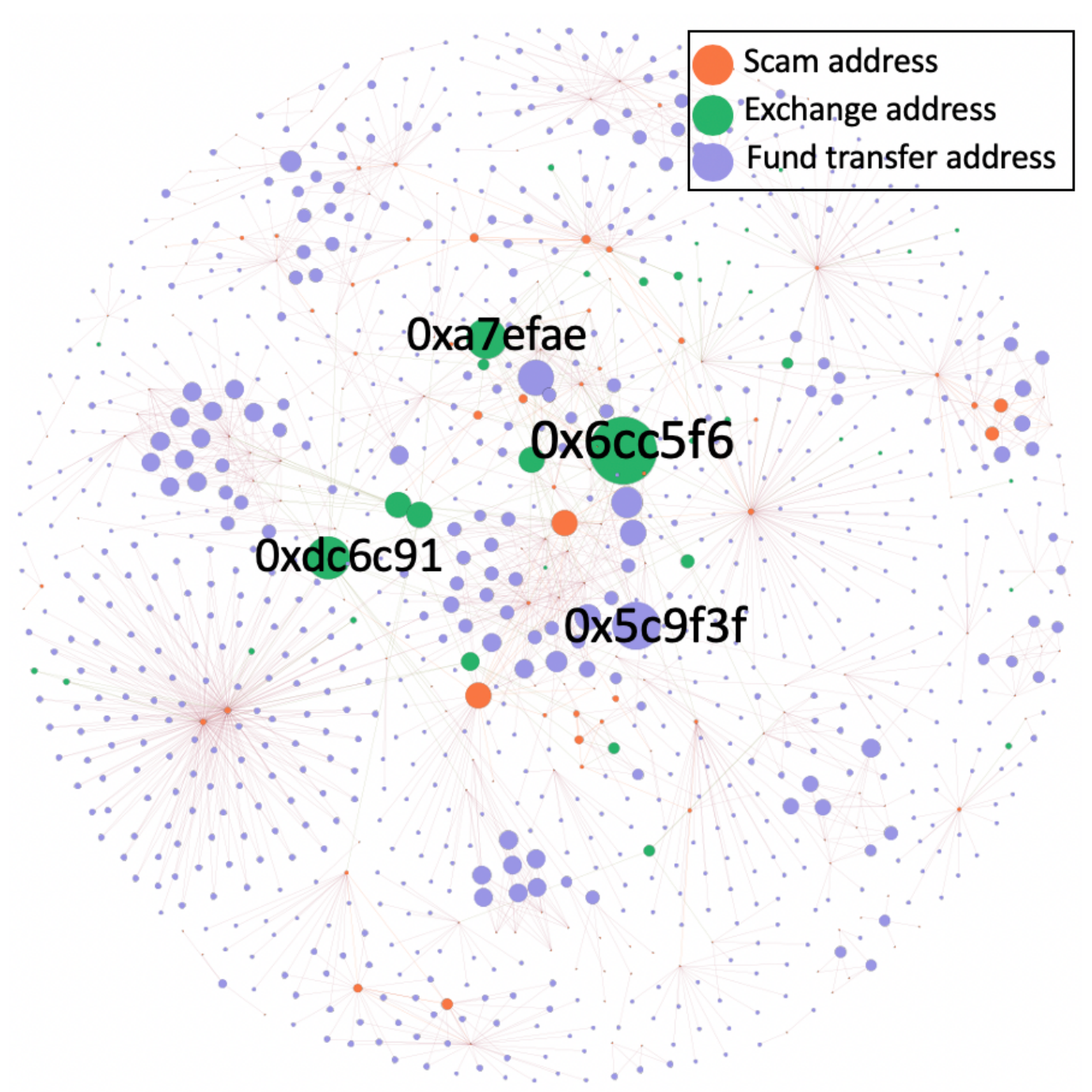}

\caption{The money flow of scam addresses (sampled 180 scam addresses).}
\label{fig:money_flow} 
\end{figure}

Next, we track the money flow of scam addresses. As an example, Figure~\ref{fig:money_flow} shows the money flow of 180 randomly selected scam addresses.
Note that, we have studied all the scam addresses indeed.
There are three types of addresses in the graph:
1) \textit{the identified scam address} (shown in orange);
2) \textit{the fund transfer address} (shown in purple), which is served as the money laundering channel, i.e., receive money from scam addresses and help the attackers transfer the money they have scammed; and
3) \textit{the exchange address} that belongs to certain known cryptocurrency exchanges. 
Note that we have purchased a premium service from an anonymous leading blockchain company to label whether an address belongs to an exchange or not.
Each edge represents the direct or indirect relationship between the addresses. 
There are 3,351 fund transfer addresses and 44 exchange addresses related to these 180 scam addresses.
\textit{Obviously, the scam addresses have transferred the money via a number of fund transfer addresses to exchanges finally.} This observation is inline with all the scam addresses.
For example, 0x6cC5F6\footnote{0x6cC5F688a315f3dC28A7781717a9A798a59fDA7b}, the exchange address of OKEx, has received ETH from 27 scam addresses we identified.
The exchange address of Tokenlon, 0xdc6c91\footnote{0xdc6c91b569C98F9F6f74d90F9BEFF99FDAf4248b}, has received ETH from 17 scam addresses.

\begin{framed}
\noindent \textbf{Answer to RQ2:} 
\textit{
We have identified two types of scams related to counterfeit tokens, i.e., airdrop scam and the arbitrage scam. 
In total, 565 active counterfeit tokens have been found involved with scams. Over 7100 victims were deceived in these scams, and the overall financial loss sums to a minimum amount of \$ 17 million (74,271.7 ETH).
}
\end{framed}

\section{Advertising Channels}
\label{sec:advertising}

In this section, we seek to explore how these counterfeit tokens are advertised to trick users. 

\subsection{Approach}
To reach the potential victims, an attacker usually provides their scam address or the counterfeit token address when they advertise the scam information.
Thus, we have implemented an automated crawler, to harvest the advertising information of these counterfeit tokens by searching the address directly.
To be specific, we first feed the 2,117 counterfeit token addresses and their corresponding scam addresses (see Table~\ref{tab:scam volume}) to Google\footnote{It is different with the step in the scam report harvesting (see Section 5.1), as here we focus on the advertising information, rather than the scam reports. Besides, we further search the identified scam address to identify the advertising information.}, and collect all the related information. Note that Google also indexes information from social network platforms including Telegram, Twitter and Facebook, etc.
Thus, we did not crawl information from these social networking platforms separately.
For the crawled results, we further eliminate information irrelevant to scams. By manually browsing the collected information, we find that there are many blockchain explorers and services (e.g., Etherscan and bloxy.info) that index all the Ethereum addresses (including the scam ones). Thus, we remove the search results from these services.
For the remaining results, we further manually verify whether they are advertising information published by attackers.

\begin{table}[t]
\caption{A Summary of the advertising channels of counterfeit tokens.}
\label{tab:scam advertising channels}
\begin{tabular}{@{}lrrr@{}}
\toprule
  Channel    & Arbitrage & Airdrop & Sum \\ \midrule
Telegraph Page  & 589       & 4       & 593 \\
Telegram Group & 52        & 76      & 128 \\ 
Twitter Page   & 0         & 25      & 25  \\ 
Facebook Page  & 0         & 22      & 22  \\ 
YouTube        & 0         & 5       & 5   \\ 
Others         & 20        & 142     & 162 \\ \midrule
Sum of Pages   & 661       & 274    & 935 \\ 
Channel Type   & 11        & 96      & 103 \\ \bottomrule
\end{tabular}
\end{table}

\subsection{The Advertising Platforms}

\subsubsection{Overall Results.} 
\begin{figure*} [t]
\centering
  \includegraphics[width=0.95\textwidth]{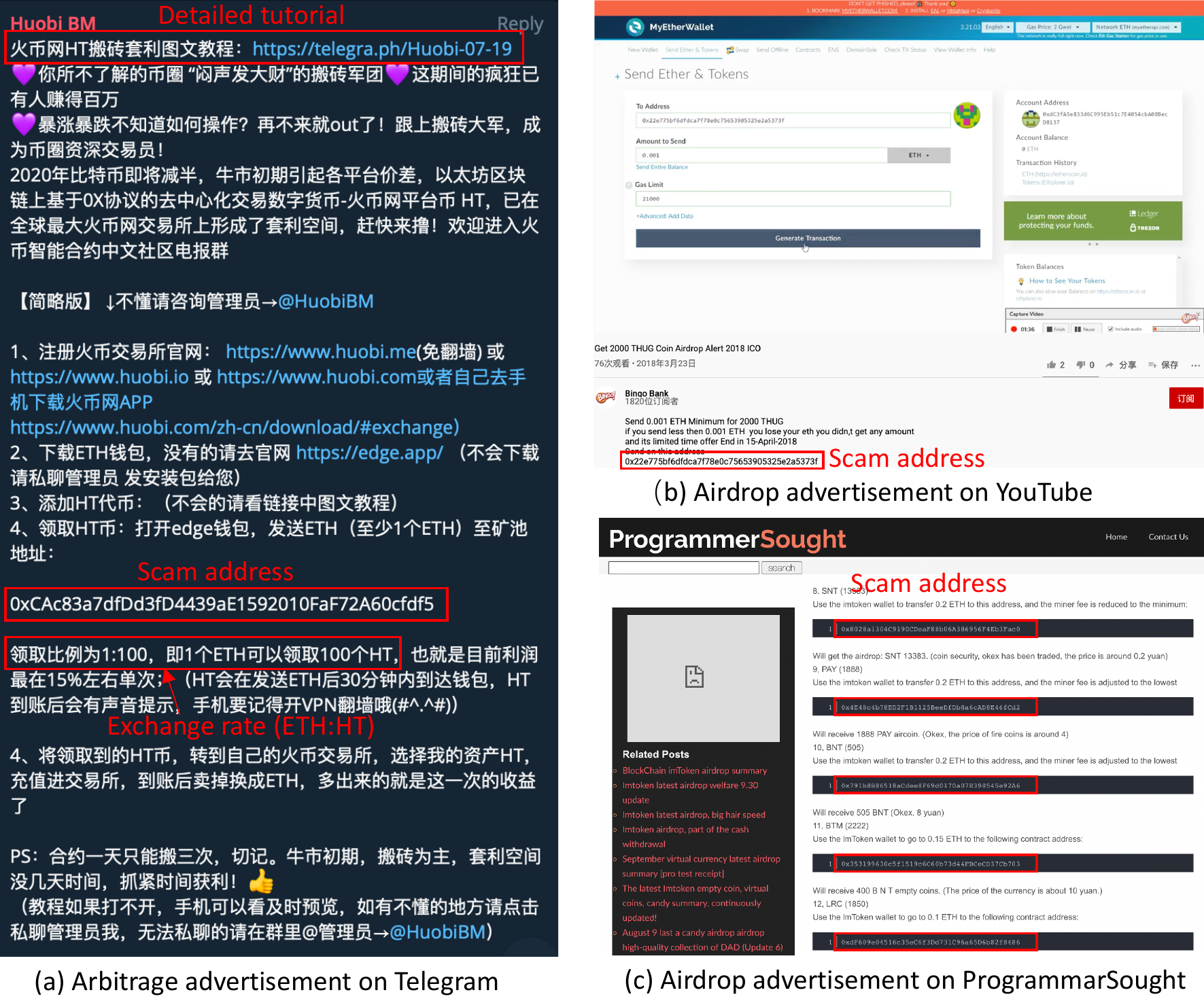}
\caption{Examples of Counterfeit Token Scam Advertising Channels}
\label{fig:advertising channels} 
\end{figure*}

We have identified 935 pieces of advertising information from 103 advertising platforms, most of which are reputable ones, including Telegraph~\cite{Telegraph}, Telegram~\cite{telegram}, Facebook~\cite{Facebook}, V2EX~\cite{V2EX}, Bctalk~\cite{Bctalk}, Bytechats~\cite{Bytechats}, Bitcointalk~\cite{bitcointalk}, Telemetr~\cite{Telemetr}, Sina~\cite{Sina}, Twipu~\cite{Twipu}, Youtube~\cite{YouTube}, Steemkr~\cite{Steemkr}, Zhihu~\cite{Zhihu}, Tgchannels~\cite{Tgchannels}, Medium~\cite{Medium}, etc.
Indeed, social networking platforms and blockchain forums are the most widely used advertising channels. 
Table~\ref{tab:scam advertising channels} summarizes the results. 
For example, we have identified 593 Telegraph pages and 128 Telegram Groups involved in the counterfeit token scams.
Obviously, airdrop and arbitrage scams have shown different advertising strategies. Although we have identified a large number (661) of arbitrage scam advertising information, they are only active on 11 platforms. However, we have identified 96 advertising channels exploited by airdrop scams.
Next, we use some case studies (see Figure~\ref{fig:advertising channels}) to illustrate how the attackers use social engineering techniques to trick users.

\subsubsection{Telegram}
Due to the anonymity of Telegram, attackers most often use Telegram groups as advertising channels. 
Here, we take the HuobiToken (HT) arbitrage scam as an example  (see Figure~\ref{fig:advertising channels} (a)) to summarize how the Telegram groups are exploited. \textbf{1) Group name and icon.} The arbitrage group always pretends to be an official group, so its name is usually like ``Huobi Official Arbitrage Tutorial Group'', and it uses the HuobiToken (HT) official icon.
\textbf{2) Advertising proposal.} The \textit{pinned message} in the Telegram group is used to publish the (fraudulent) arbitrage tutorial. The tutorial explains in detail the workflow of arbitrage and the popularity of arbitrage tokens. 
Besides, scammers will insert official and reputable website links in the tutorial, which increases its credibility.
\textbf{3) Customer Service.} 
Every Telegram group has a fake official customer service. 
The customer service also imitates the real HuobiToken (HT) telegram official group. As aforementioned, their main purpose is to lead the victim to the secondary scam. Besides, once the victim found out that they have been deceived, the fake customer service will kick out those who expose the scam from the Telegram group.
\textbf{4) Bot Accounts.} For the scam Telegram groups we identified that every group has thousands of members, but most of them are robots. The robots in the group falsify the conversation, making victims mistakenly believe that the arbitrage is real and profitable.

\subsubsection{YouTube}
The content sharing platforms can also be exploited to distribute scam information.
We use scams on YouTube to show the advertising process (see Figure~\ref{fig:advertising channels} (b)).
They usually pretend to be the official tutorial of airdrop, which illustrate the complete airdrop process for the potential victims.
To increase their credibility, 1) they show some fake information on the video which states that many users have received tokens; 2) they always imitate famous people (e.g., Bill Gates, and Vitalik Buterin, the founder of Ethereum); and 3) official websites are embedded in the video.

\subsubsection{Blockchain Forums}
Online forums are also the main targets for attackers to reach potential victims, mostly for airdrop scams, based on our observations. 
We take the airdrop scam we found from ProgrammarSought as an example (see Figure~\ref{fig:advertising channels} (c)).
The article published by the attacker is also a tutorial of the airdrop. To increase its credibility, 1) it provides a list of airdrop tokens, which might be mixed with real official airdrops. Some of them are free airdrops, thus victims can get tokens for free, which makes victims believe the credibility of the listed other (scam) airdrops. 2) it also shows the screenshots of tokens received after the airdrop is completed.

\begin{framed}
\noindent \textbf{Answer to RQ3:} 
\textit{
A number of reputable platforms have been abused by attackers to help spread fraudulent information on counterfeit tokens. 
Social network platforms like Telegram, Twitter and Facebook are the main targets of attackers. 
Various social engineering techniques have been adopted to trick users.
}
\end{framed}

\section{Discussion}

\subsection{Implication}
Our observations are of key importance to stakeholders in the community.

\textbf{The governance of the cryptocurrency.}
Considering the large number of counterfeit tokens and scams we discover, the governance of cryptocurrency needs to be improved. There is a need to design policies to regulate cryptocurrency naming schemes.
However, like domain squatting issues that remain in the wild for years, it is not easy to fully address counterfeit token issues. Even if Ethereum disallows creators to release ERC-20 tokens with the same identifier names, a number of other kinds of attacks in the domain-squatting field, including typosquatting (e.g., ``yuotube.com''), bit-squatting (accidental random bit flip,
e.g., ``yo5tube.com''), homograph-based squatting (e.g., ``y0utube.com''
), and sound squatting (e.g., ``yewtube.com
'') would be easily applied to the token names/symbols to mislead users. 

\textbf{Cryptocurrency wallets, exchanges and blockchain browsers.}
Cryptocurrency wallets, exchanges and blockchain browsers have the responsibility to detect counterfeit tokens and protect users from being scammed. Our approach can be be integrated within major exchanges and wallets to stop such scams. For example, once a new counterfeit token is found, our approach can help flag all the suspicious scam addresses related to it and warn users before they interact with these addresses. We observe that major blockchain browsers (e.g., Etherscan and Bloxy) have started to flag scam addresses using their own approaches, thus our work can also be implemented in these browsers to help flag counterfeit token related scams and remind users timely.

\textbf{Cryptocurrency creators}
The official cryptocurrency creators should be aware of the counterfeit token abuse. They should take the responsibility to search and identify counterfeit tokens and even fake social networking accounts (e.g., Telegram and Twitter). In such cases, cryptocurrency creators could then take actions to mitigate possible abuses (e.g., by reporting them to regulators and investors). Furthermore, they should regularly post public announcement to remind users.

\textbf{Investors}
Awareness should also be raised among investors. For educational purposes, we commit to post regular tutorials and reports to provide a means for regulators, cryptocurrency creators and investors to learn more about counterfeit tokens.
More importantly, investors should keep in mind that there is no such thing as a free lunch in the cryptocurrency world.

\textbf{Advertising Channels.}
Finally, as we have identified a number of advertising channels exploited by attackers, including reputable ones, it is also urgent to regulate the contents published on these platforms, which can help reduce the propagation of scams.

\subsection{Limitation}
Our work carries several limitations.
First, we only study counterfeit tokens related to the top-100 official cryptocurrencies on Ethereum. 
Although our observation suggests that the attackers are more likely to target popular tokens with a high market capitalization rank, it is quite possible that there are some counterfeit tokens targeting other cryptocurrencies beyond our study.
Second, counterfeit cryptocurrencies can target official tokens on any cryptocurrency platforms, while we only focus on the counterfeit ERC-20 tokens on Ethereum, as ERC-20 is the most popular token standard,  accounting for over 90\% of alternative tokens in the blockchain ecosystem. Nevertheless, we agree that the counterfeit token might exist in other blockchain platforms like EOSIO and Tron. We leave them for  future work.
Third, we have characterized two typical scams related to the counterfeit tokens by resorting to existing scam reports. However, it is possible that there are other scams related to counterfeit tokens we did not cover.
Finally, although we tried our best to understand the overall workflow of counterfeit cryptocurrency scams, we lack a deep understanding of the attackers behind the scams. The social engineering techniques and the advertising posts we identified may be just the tip of the iceberg. Nevertheless, this paper presents the lower bound impact analysis of the counterfeit cryptocurrency scams.

\section{Related Work}

\subsection{Blockchain Scams}
Since the birth of blockchain, various kinds of scams have emerged. 
A number of studies have characterized blockchain scams. 
Vasek and Moore~\cite{vasek2015there} surveyed the presence of Bitcoin scams, including Ponzi schemes, mining scams, scam wallets, and fraudulent exchanges. 
After that, some other studies have characterized various scams including Ponzi schemes~\cite{chen2018detecting, bartoletti2020dissecting, bartoletti2018data, vasek2018analyzing, chen2019detecting, toyoda2017identification, toyoda2019novel}, scam Initial Coin Offerings (ICOs)~\cite{milind2020future, liebau2019crypto, daniel2019cryptocurrencies, zetzsche2017ico}, market manipulation of cryptocurrencies~\cite{gandal2018price, chen2019market, hamrick2018economics, chen2019detecting, hamrick2018examination}, blockchain honeypots~\cite{torres2019art}, and phishing scams~\cite{wu2019phishers, phillips2020tracing,xia2020dont}.

ICO scams are most relevant to this paper. For example,Alexander et al.~\cite{alexander2020predicting} built a predictive model by applying natural language processing (NLP) and machine learning techniques to detect ICO scams.
Shuqing et al.~\cite{bian2018icorating} created ICORating, a learning-based cryptocurrency rating system. They have analyzed 2,251 cryptocurrencies from a number of perspectives, including whitepapers, founding teams, Github repositories, websites, etc. 
\textit{Counterfeit cryptocurrency, as a new emerging threat, has not been systematically studied yet}. In this work, we take the first step to characterize counterfeit tokens and study their relevant scams.

\subsection{ERC-20 Tokens}

A few studies have characterized the ERC-20 token ecosystem~\cite{chen2020traveling, FriedhelmERC20, morales2020user, dyson2020scenario, tokenscope}. For example, Chen et al.~\cite{chen2020traveling} investigated the Ethereum ERC20 token ecosystem to characterize the token creator, holder, and transfer activity. 
Friedhelm Victor et al.~\cite{FriedhelmERC20} provided an overview of more than 64,000 ERC20 token networks and analyzed the top 1,000 from a graph perspective.
Besides, there are some studies dedicated to optimizing the ERC-20 token standard~\cite{mayer2020batpay,rahimian2019resolving} or using ERC-20 token contract to address practical issues~\cite{christodoulou2020decentralized,koscina2019plasticcoin}. For example, Mayer et al.~\cite{mayer2020batpay} proposed a proxy scaling solution for ERC-20 tokens named BatPay, which is suitable for micropayments in one-to-many and few-to-many scenarios and can reduce gas cost of transactions. Christodoulou et al.~\cite{christodoulou2020decentralized} designed a smart contract that can interact with any ERC-20 token to help decentralised organizations run public voting campaigns and engage token holders in voting decision.

\subsection{Blockchain Transaction Analysis}

A number of studies have investigated
blockchain systems by performing transaction-based analyses.
Several studies are focused on Bitcoin~\cite{FergalSC2011, DoritFC13, alex2014deanonymisation, ChenDF15, MichaelArxiv15, DamianoCN16, SilivanxayICDMW18}, including de-anonymization and money laundering detection, by using graph-based approaches.
Researchers have also investigated Ethereum and EOSIO by using transaction-based analyses.
For example, Chen et al.~\cite{TingINFOCOM18} performed a graph-based analysis of Ethereum to characterize activities including money transfer, smart contract creation and smart contract invocation.
Huang et al.~\cite{EOSIOStudy} analyzed the transactions on EOSIO blockchain, and developed techniques to automatically detect bots and fraudulent activities.

\section{Conclusion}

This paper has presented the first in-depth measurement study of counterfeit tokens on Ethereum. Our study has revealed that counterfeit tokens are prevalent in the cryptocurrency ecosystem, thereby motivating the need for more efforts to identify and prevent cryptocurrency abuses. 
We have characterized two kinds of scams related to counterfeit tokens, and designed methods to identify airdrop scams and arbitrage scams. At least 7,104 victims have been scammed and the overall profit is over \$$17,352,840.00$. By studying the advertising channels for the counterfeit tokens and scams, we find 103 platforms have been exploited to spread fraudulent information.

\section*{Acknowledgment}

We sincerely thank our shepherd Prof. Stefan Schmid (University of Vienna) and all the anonymous reviewers for their valuable suggestions and comments to improve this paper. This work was supported by the National Natural Science Foundation of China (grants No.61702045 and No.62072046), Hong Kong RGC Project (No. 152193/19E), the Fundamental Research Funds for the Central Universities, Leading Innovative and Entrepreneur Team Introduction Program of Zhejiang (2018R01005). Haoyu Wang (haoyuwang@bupt.edu.cn) is the corresponding author.

\balance

\bibliographystyle{plain}
\bibliography{cite}


\end{document}